\documentclass[onecolumn]{aastex61}

\usepackage{amsmath}
\usepackage{graphicx}

\usepackage{epsfig}

\usepackage{wasysym}
\usepackage{lipsum}
\usepackage{mathrsfs}
\usepackage{float}
\usepackage{rotating}
\usepackage{graphicx}
\usepackage{epstopdf}
\usepackage{array}
\newcolumntype{P}[1]{>{\centering\arraybackslash}p{#1}}
\newcolumntype{M}[1]{>{\centering\arraybackslash}m{#1}}
\newcounter{chem}
\newcounter{temp}

\usepackage{mhchem}

\providecommand{\e}[1]{\ensuremath{\times 10^{#1}}}

\usepackage{color}

\usepackage{gensymb}

\usepackage{filecontents}

\begin{document}

\shorttitle{Volatile Content on Young Earths}
\shortauthors{Chen, H. \& Jacobson, S. A.}

\title{{\bf Impact Induced Atmosphere-Mantle Exchange Sets the Volatile Elemental Ratios on Primitive Earths}}

\author[0000-0003-1995-1351]{Howard Chen}
\affil{NASA Goddard Space Flight Center, Planetary Environments Laboratory,  Greenbelt, MD 20771, USA }
\affil{GSFC Sellers Exoplanet Environments Collaboration}
\affil{Oak Ridge Associated Universities}

\author[0000-0002-4952-9007]{Seth A. Jacobson}
\affil{Department of Earth and Environmental Sciences, Michigan State University, East Lansing, MI 48824}

\correspondingauthor{Howard Chen} 

\begin{abstract}
Conventional planet formation theory suggests that chondritic materials have delivered crucial atmospheric and hydrospheric elements such as carbon (C), nitrogen (N), and hydrogen (H) onto primitive Earth. However, recent measurements highlight the significant elemental ratio discrepancies between terrestrial parent bodies and the supposed planet building blocks. Here we present a volatile evolution model during the assembly of Earth and Earth-like planets. Our model includes impact losses, atmosphere-mantle exchange, and time dependent effects of accretion and outgassing calculated from dynamical modeling outcomes. Exploring a wide range of planetesimal properties (i.e., size and composition) as well as impact history informed by N-body accretion  simulations, we find that while the degree of CNH fractionation has inherent stochasticity, the evolution of  C/N and C/H ratios can be traced back to certain properties of the protoplanet and projectiles. Interestingly, the majority of our Earth-like planets acquire superchondritic final C/N ratios, implying that the volatile elemental ratios on terrestrial planets are driven by the complex interplay between delivery, atmospheric ablation, and mantle degassing.
\end{abstract}

\keywords{astrobiology -- planets and satellites: atmospheres --  planets and satellites: terrestrial planets}

\email{howard@earth.northwestern.edu}

\section{Introduction} 
\label{sec:intro} 

The timing and sources for the acquisition of Earth's major volatiles from pre-Solar nebular gases laid the foundation for the planet's geology, atmosphere, and habitability \citep{kasting+03,MartyEt2012EPSL}. While Earth itself was accreted by materials from a mixture of chondritic sources and planetary precursors \citep{Drake+Righter2012NATURE,MartyEt2016EPSL,MorbidelliEt2000EPSL}, Earth's abundance and composition are incompatible with the originally volatile-rich disk materials from which it was built \citep{halliday+13}, suggesting that these materials have undergone significant thermal alteration, transfer, and/or differentiation prior to or during accretion. One or a combination of these processes are needed to explain the depletion of Earth's noble gas abundances (by a factor of 40 relative to CI carbonaceous chondrites), and the depletion C and N (by a factor of ${\sim}1000$; \citealt{lodders+03}). Hence how Earth has acquire and retained key moderately volatiles (C, N, H) and extremely volatile (noble gas) ingredients during planet formation is a complex problem involving interactions between the impactors and the parent body, as well as between the their principal reservoirs (i.e, mantle, crust, and fluid envelopes). 
%Models based on meteoritic data conclude that bodies originating near Earth's current orbit are typically dry and devoid of volatiles (particularly the enstatite chondrites; \citealt{Drake+Righter2012NATURE}), which is severely at odds with the properties of the inner terrestrial planets.
%The processes involved in the delivery, loss, transfer, and timing of volatiles and their effects on the present-day atmospheres/hydrospheres is therefore of paramount concern. Efficiency of this process is complicated and depends on the dynamics of the early Solar System, sizes and compositions of the accreted bodies, the state of the underlying interior. 

The formation of Earth and the terrestrial planets of our Solar Sytem involves the collection of ${\sim} 10^{12}$ meter- and kilometer-sized building blocks on the order of ${\sim}40-100$ Myr \citep{Chambers2004EPSL,jacobson+14}.  Strong radial mixing during late-stage accretion is predicted when the local density of planetesimals and embryos becomes comparable \citep{Chambers2001Icar,OBrienEt2006Icar,RaymondEt2009Icar}.
During these intense mixing episodes, proto-planets accreted scattered embryos and planetesimals and the feeding zones can reach up to several AUs \citep{RaymondEt2006Icar}. The accretion of these planetesimals led to the development of embryos that are comparable to their isolation masses \citep{ida+makino93,hayashi+81}. Embryo merging events via giant giants eventually formed Earth and Venus and potentially contributed to the volatile inventories of both planets. However, isotopic constraints show that that Earth's water matches carbonaceous chondrites, which originate from the outer asteroid belt and further out \citep{MorbidelliEt2000EPSL}. In order for Earth to have obtained its present volatile (i.e., water) abundances, planetary embryos must thought to have experienced intense bombardment by carbon and water-rich CI chondrite-like bodies from the outer asteroid region early in their evolution stages.
As such, the final composition of the inner planets is heavily dependant on the prescribed initial composition distribution (as well as the configuration of Jupiter and Saturn) \citep{Chambers2001Icar,OBrienEt2006Icar,RaymondEt2009Icar,WalshEt2011NATURE}. 
Furthermore, if scattering by the inward and outward migration of Jupiter is efficient (e.g., the Grand Tack Paradigm), then a number of volatile-rich asteroids from the trans-Saturn region could have collided with the inner terrestrial planets \citep{OBrienEt2014Icar}. Thus to understand the consequence of both both impact loss and delivery as well as volatile transfer across sub components of the BSE and the core, a time-marching model that includes stochastic accretion and feedback effects between reservoirs is needed.

The accretion of atmophile elements (i.e., gas-phase species typically found in the atmosphere) is not a simple summation of individual building block materials. Escape and exchange mechanisms between and within the interacting/colliding bodies can be non-trivial and understanding them requires experimental constraints and theoretical models of accretion and differentiation processes. For instance, high energy photon driven hydrodynamic escape is thought to have removed light atmospheric species such as H and He in the ancient Venusian atmosphere (e.g., \citealt{Kasting+Pollack1983Icar,ErkeavEt2014}), resulting in the observed isotopic fractionation \citep{Grinspoon1987,Donahue1982}, and could even affect the proto-atmospheres of terrestrial and gaseous exoplanets\footnote{Atmospheric escape caused by non-thermal processes such as solar flaring and energetic particles precipitation are observed in elsewhere \citep{Louca+2022arXiv,howard+2022} and could also be important in the early solar system, addition to hydrodynamic and impact losses.} \citep{howe+2020,Rogers+2021}.
Impacts can also drive atmospheric loss to space; all major planetary bodies in the Solar System have suffered intense bombardments during and after accretion \citep{SchaberEt1992JGR,Albar2009,AhrensEt1993,schlichting+18}. Core forming metals could have removed a substantial fraction of carbon from the bulk silicate Earth (BSE) \citep{wood+06}. Although some studies suggest that N could be soluble in the core (e.g., \citealt{Grewal+21}), this does not necessarily mean that N will actually be chemically partitioned there.%, which might preceded and the 

%paragraph on previous work such as, that used a simple single stage whole model of a magma ocean equilibrrating wtih an overlying atmosphere and lower transition to the core. say something about solubilities and partition coefficients. say why these static models are insufficient as they do not consider time dependent effects such as competition between volatile delivery and core sequestration. nd should hence be studied with coupled time-resolved modeling tools.

Astrophysical models of solar system and extrasolar planets  typically treat the planet with a single volatile reservoir  \citep{ValenciaEt2010A&A,Chen+Rogers2016ApJ}.
On terrestrial planets however, extremely volatile (e,.g., true atmophiles such as noble gases) and moderately volatiles (C, N, H, S) can reside in multiple reservoirs--the atmosphere (or the hydrosphere), the mantle, and even the core, as many of them are strong lithophiles and siderophiles. Surface interactions with the primordial liquid mantle (e.g., degassing and ingassing) have also shown to be important to the thermal evolution and retention of terrestrial atmospheres \citep{avice+20,grewal+20}. As terrestrial planets formed as molten or partially molten states \citep{Solomon1979} the vertical distribution of volatiles between the shallow mantle and atmosphere hold clues to the initial phases of planetary growth \citep{ZahnleEt1988Icar, Elkins2008EPSL}.
Further, the development of thick steam atmospheres, expected during large impacts, could lead to strengthened ingassing of volatiles into the deep interior of nascent Earth and/or the formation of global-scale primordial oceans \citep{Matsui+Abe1986,ZahnleEt1988Icar}.
%This suggest that modeling mantle-atmosphere exchanges is important to more realistically account for atmospheric accretion.

Understanding the volatile abundance and composition on proto-worlds will require understanding of not only the properties of the impactors but also the effect of each impact on the planet embryo's mantle and atmosphere. Due to the chaotic nature of planetary interactions however, the cumulative volatile delivery to a planet varies with the initial configuration and mass distribution of the system. To account for planetesimal accretion, the usual approach is to assume a constant impact flux based on inferred impact history (e.g., \citealt{Matsui+Abe1986,ZahnleEt1988Icar}). While such a ``fixed-flux" approach has been widely applied, they neglect the stochasticity inherent to accretion processes. N-body planetary accretion simulations demonstrate that with similar initial conditions, two simulations can result in very different accretion histories  \citep{Chambers2001Icar}. More recent work has applied a statistical model of stochastic bombardment to study the formation of Earth's proto-atmosphere during late veneer (1-100 Myr; \citealt{sinclair+20}), and application to earlier epochs ($100 - 10^6$yr) during the earlier stages of planetary formation.

To date, only a handful of studies attempted to integrate mantle chemistry calculations and N-body simulations (e.g., \citet{rubie+15}) and to our knowledge, no study has integrated atmosphere-mantle exchange with N-body dynamical outcomes. Particularly, {\it i)} modern models of core-mantle, metal-silicate differentiation model do not include impact escape and, {\it ii)} studies of atmospheric escape do not interact with chemical evolution and account for stochastic impact events derived from N-body accretion scenarios. Importantly, as the consequence of each impact depends on the evolution of the underlying volatile inventory and the state of the overlain atmosphere,  a model that incorporates coupled/interactive effects is needed. In this paper, we make such a contribution by introducing an atmosphere-mantle volatile evolution model of nascent Earth.

\subsection{The Chemical Composition of Nascent Earth}

Today, the cycling of major volatile species such as carbon dioxide, nitrogen, and water through our planet's interior and surface strongly influences  its present-day geochemistry, long-term surface climate, and the stability of the biosphere \citep{mcgovern+89,sleep+01a}. Carbon concentration in the mantle is estimated to be 2$\times 10^{19}$ kg \citep{cartigny+08}. The global nitrogen budget is estimated to be 8$\times 10^{18}$ kg \citep{goldblatt+09}\footnote{Estimates for the carbon content of the mantle plus atmosphere can vary by a factor of 10 (and nitrogen by ${\sim}2\times$), see for example \citet{halliday+13} and references therein}. Finally, the estimated mantle water content is about 200 ppm, but \citet{MartyEt2012EPSL} have posited that up to 90\% of water are contained in the mantle which may contain at least an ocean mass worth of H$_2$O. In addition to absolute concentrations, elemental ratios (e.g., C/N, C/H, C/S  \citealt{hirschmann2016am}) can offer insights into fractionation events because they are often associated with specific flux pathways and cosmochemical sources that hold clues to their origin and history. 

The C/N values of the interstellar medium, the early nebula, comets, and various types of chondrites differ substantially (i.e., from ${\sim} 5$ to ${\sim} 200$), suggesting that the initial nebular imprint, differentiation, thermal metamorphism, outgassing, and atmospheric escape could have all influenced the eventual C/N of the parent body \citep{MartyEt2012EPSL,dasgupta2013gca,chi2014gca,tucker+14}. In the bulk silicate Earth (BSE), nitrogen and carbon are depleted relative to other major volatiles  \citep{MartyEt2012EPSL,bergin2015pnas}, as well as to extreme volatiles such as noble gases \citep{marty+20}. In particular, the modern BSE is found to have superchondritic C/N, which is counter-intuitive as core formation should substantially reduce the BSE's C/N due to the more siderophile nature of C \citep{dalou+17}. Note that H is much less siderophilic compared to C and N, which could plausibly explain the subchondritic C/H ratio in the BSE \citep{dasgupta2013gca}. Lastly, recent work suggest that the C/S value in the BSE could be the result of open-system silicate melting and  carbon loss \citep{hirschmann+21}. Any satisfactory mechanism must not only simultaneously explain all three degrees of elemental fractionation, but also involve the differential removal of C, N, and H relative to the noble gases \citep{halliday+13}. 
%While core formation should reduce carbon over nitrogen abundances, the loss of an early nitrogen-dominated atmosphere could produce the observed superchondritic C/N \citep{chi2014gca,tucker2014epsl}. Due to due to the more siderophilic nature of carbon,  \citet{hirschmann2016am} suggested that materials delivered to the primitive Earth should already have enhanced C/N.  More recently, \citet{grewal2019science} used a very fined tune scenario of a Moon-forming impactor to explain the observed C/N. However, a more comprehensive assessment  accounting for time-dependent effects and feedbacks between mantle and atmospheric exchanges, in addition to incorporating stochastic delivery,   is warranted.

\section{Model Setup}
\label{sec:method}
We built an analytical model of atmophile accretion to compute the volatile species evolution as a function of growing Earth-analogs.
Integrated with N-body dynamical simulations, our model includes three reservoirs (atmosphere, mantle, core) and calculates the temporal evolution of C, N, and H amounts as a function of the growing planet.
The model accounts for impact degassing/ingassing, atmospheric ablation, and core-mantle-atmosphere equilibration. 
By modeling how these processes affect the compositions of the interacting bodies for every impact, we track and store the changing volatile content in the reservoirs as the accretion advances. When the online version of the paper is published, the model package will publicly available at \url{https://github.com/hwchen1996/vol_accret}.
Model description is provided below and a more detailed one, including the mathematical formalisms, can be found in the Supplementary Materials.

\subsection{Model Processes and Components}

As the assemblage of terrestrial planets is inextricably related to the formation of the volatile inventory, a variety of source/sinks and their associated processes should ideally be considered in an integrated model.
These processes include:

\begin{itemize}
    \item initial composition of solids,
    \item dynamical evolution and impact history,
    \item metal-silicate-gas exchange, and
    \item host star activity and radiation environment.
\end{itemize}

Each component of our model design addresses one or more of the above process, with an overall goal of assessing impact-induced effects on the volatile content evolution on planetary building blocks. Volatile delivery via impacts may contribute to the global inventory by simply devolatilizing their materials onto the planetary surfaces \citep{zahnle+07}. Degassing from the deep planetary interior may also supply additional gas to the atmosphere. Ingassing and impacts induced ablation could remove significant amount gases from the atmospheric reservoir, and any species not removed could then be distributed within the planet according to their partition and solubility constants. Calculating both retention and loss processes, as well as potential coupled and feedback effects, is critical to backtrack and predict the volatile abundances on terrestrial planets. In this work, these physical processed are incorporated by modeling:

\begin{itemize}
    \item Stochastic impact history:  The outputs from a suite of N-body simulations based on the Grand Tack Paradigm \citep{WalshEt2011NATURE,jacobson+morbi14} are used by our volatile growth model. The individual embryo mass and embryo-to-planetesimal ratios were varied.
    \item Volatile composition of cosmochemical sources: Chondritic materials (enstatite chondrites, ordinary chondrites, carbonaceous chondrites) are assigned volatile fraction values based on their initial heliocentric distances. Note that while we consider only post-disk evolution, planetesimal and embryo compositions can serve as records of the volatile evolution of the gaseous nebular disk. Further, when the materials reach a 10-100 km in size, they are too small to accrete gas and they will no longer equilibrate with the surrounding materials/chemistry. 

    \item Accretion of planetesimals:  Drawing the asteroidal size distribution, accretion of super-planetesimals from the N-body simulations are broken down into smaller projectiles. These materials, resembling true planetesimals in size, are added successively onto the parent body.
    \item Atmospheric impact erosion: Effects of both giants impacts and planetesimals are approximated using the non-local escape prescription of \citet{SchlichtingEt2015Icar}.
    \item Mantle degassing and ingassing: The formation of magma oceans promotes volatile transfer between the mantle and atmosphere. We use Henrian solubilities (e.g., \citealt{hirschmann2016am}) to calculate the mass exchanged during each impact event. One important aspect of this model is that the volatiles dissolved in the mantle is determined solely by the mass of the overlying atmosphere.
    \item Core formation: During core formation, metals and highly siderophile elements such as carbon are segregated to the core. We use the prescription from \citet{deguen2011experiments} to calculate mantle-core equilibration of carbon. 
\end{itemize}

Ultimately, this model seeks to {\it (i)} document  the accretion histories of volatiles elements (N, C, H) and  their primary gaseous phases (e.g., N$_2$, CO$_2$, H$_2$O), and {\it (ii)} explicitly quantity their movements across planetary reservoirs.  With the above inclusions, the potential complexity of gain/loss events via their dependencies on the composition and timing of the accreted materials can be evaluated. 

\begin{figure*}[t] 
\begin{center}
\includegraphics[width=0.6\columnwidth]{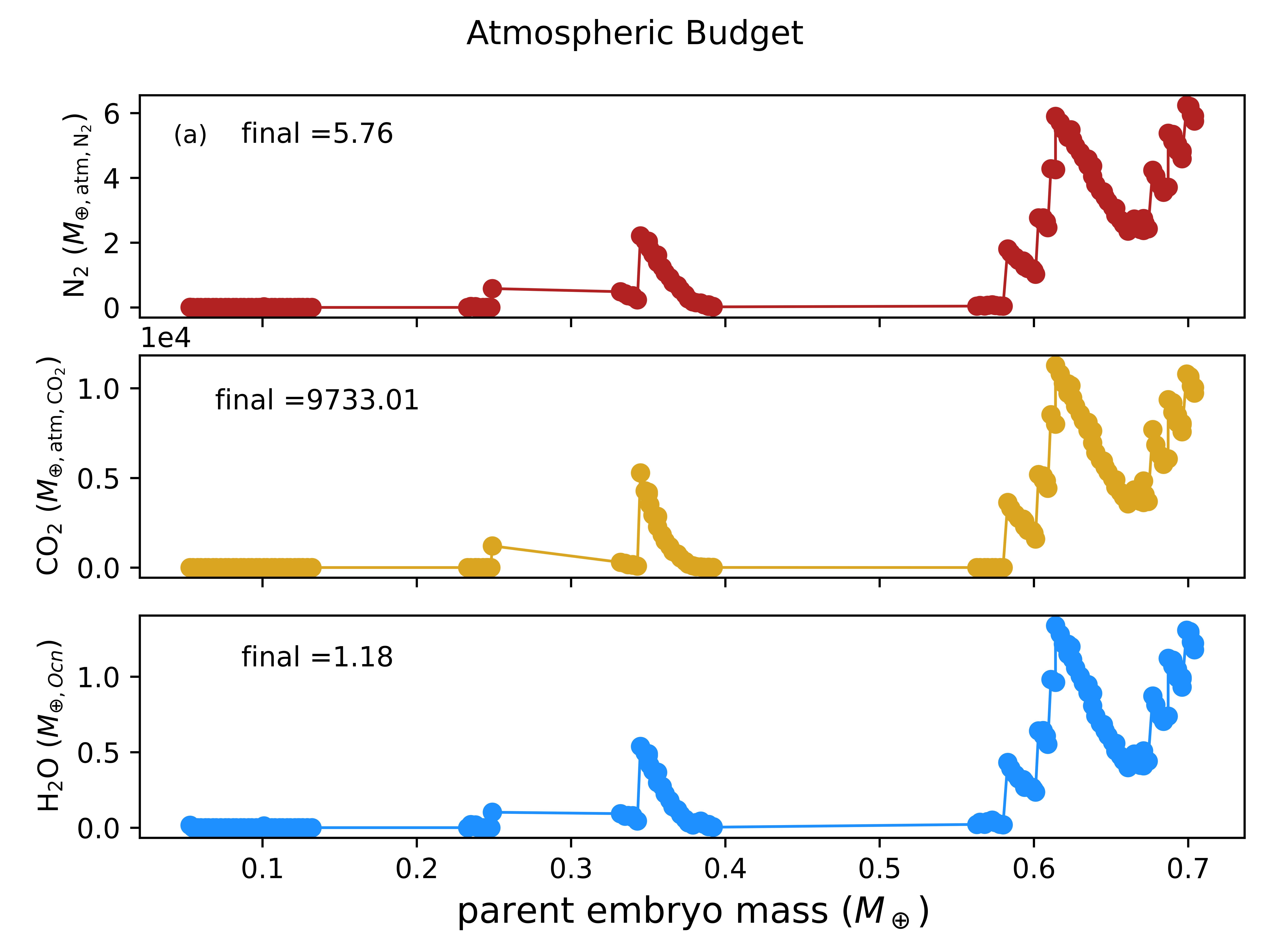}
\includegraphics[width=0.6\columnwidth]{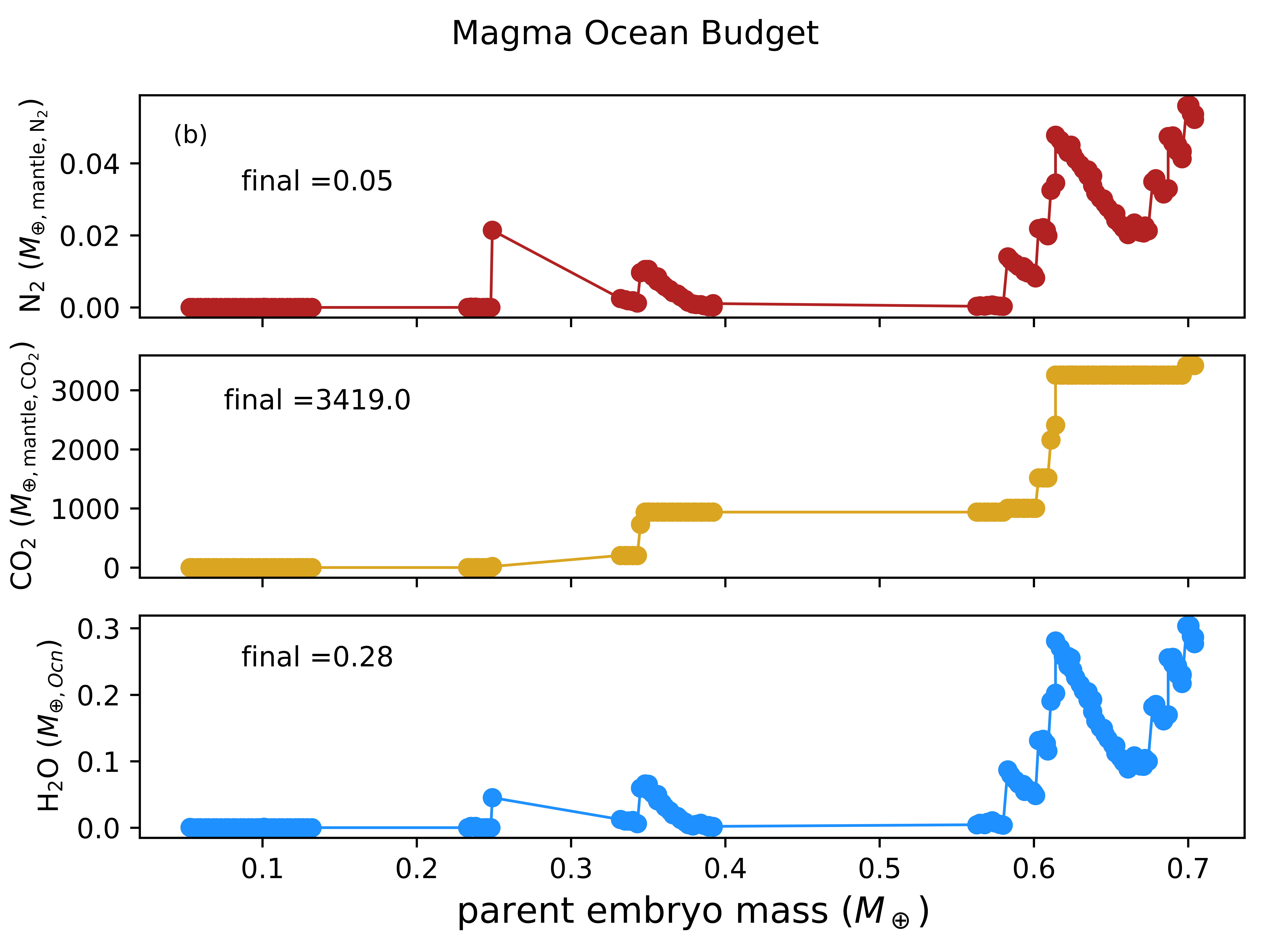}
\caption{\label{fig:evo} Time evolution of N$_2$, H$_2$O, and CO$_2$ in the atmosphere and mantle as an Earth-like planet grows, normalized by the present-day values of their respective reservoirs (i.e., atmosphere or mantle ). These results correspond to the reference case. }
\end{center}
\end{figure*} 

\begin{figure*}[t] 
\begin{center}
\includegraphics[width=0.8\columnwidth]{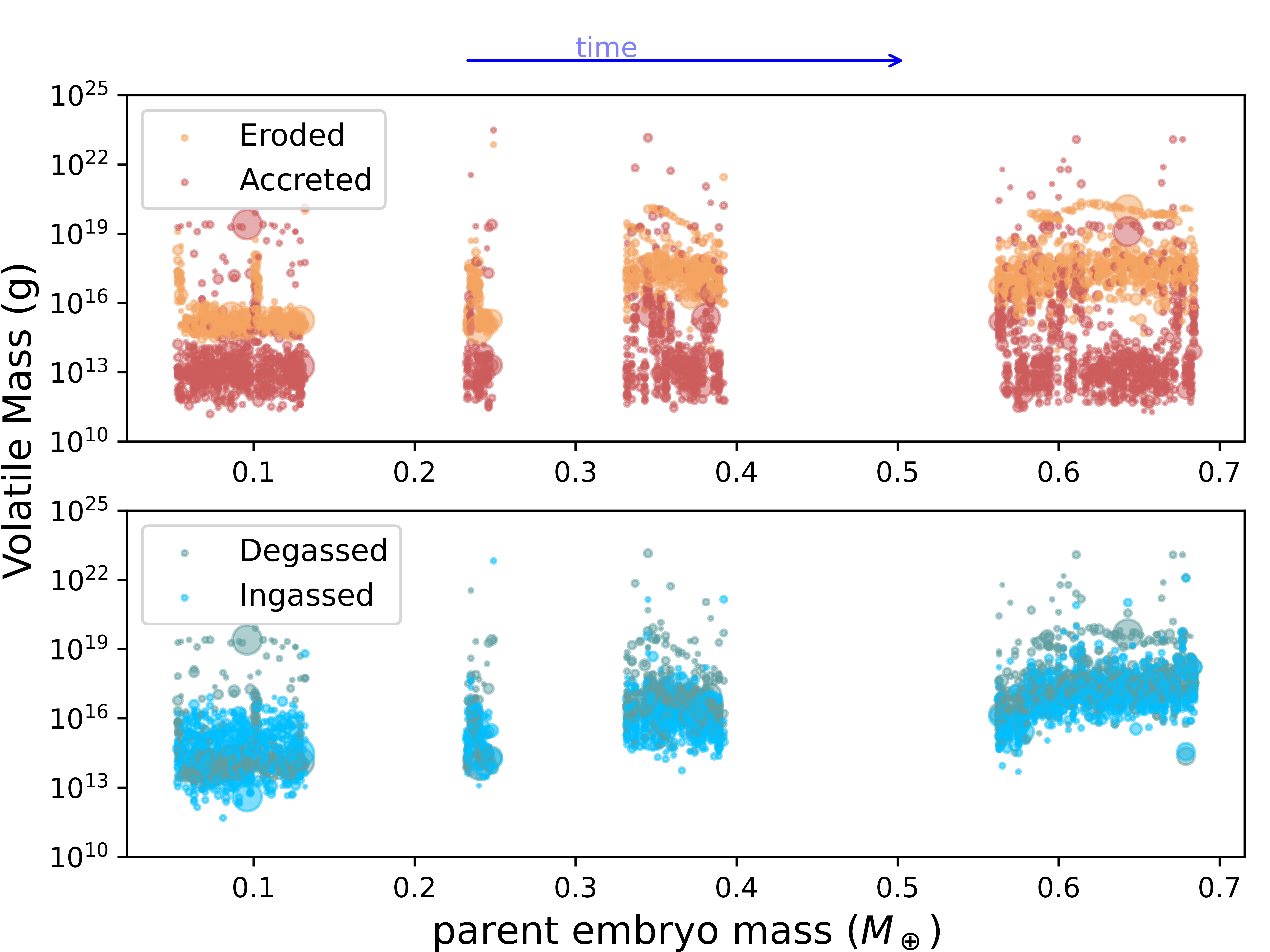}
\caption{\label{fig:esc} Volatile mass accreted/eroded (a) and mass degassed/ingassed (b) with a growing Earth. We randomly sample one out of $10^4$ planetesimals and document the volatile mass transferred during each impact event. Marker size denotes impactor radius.  }
\end{center}
\end{figure*}  

\begin{figure}[h] 
\begin{center}
\includegraphics[width=0.7\columnwidth]{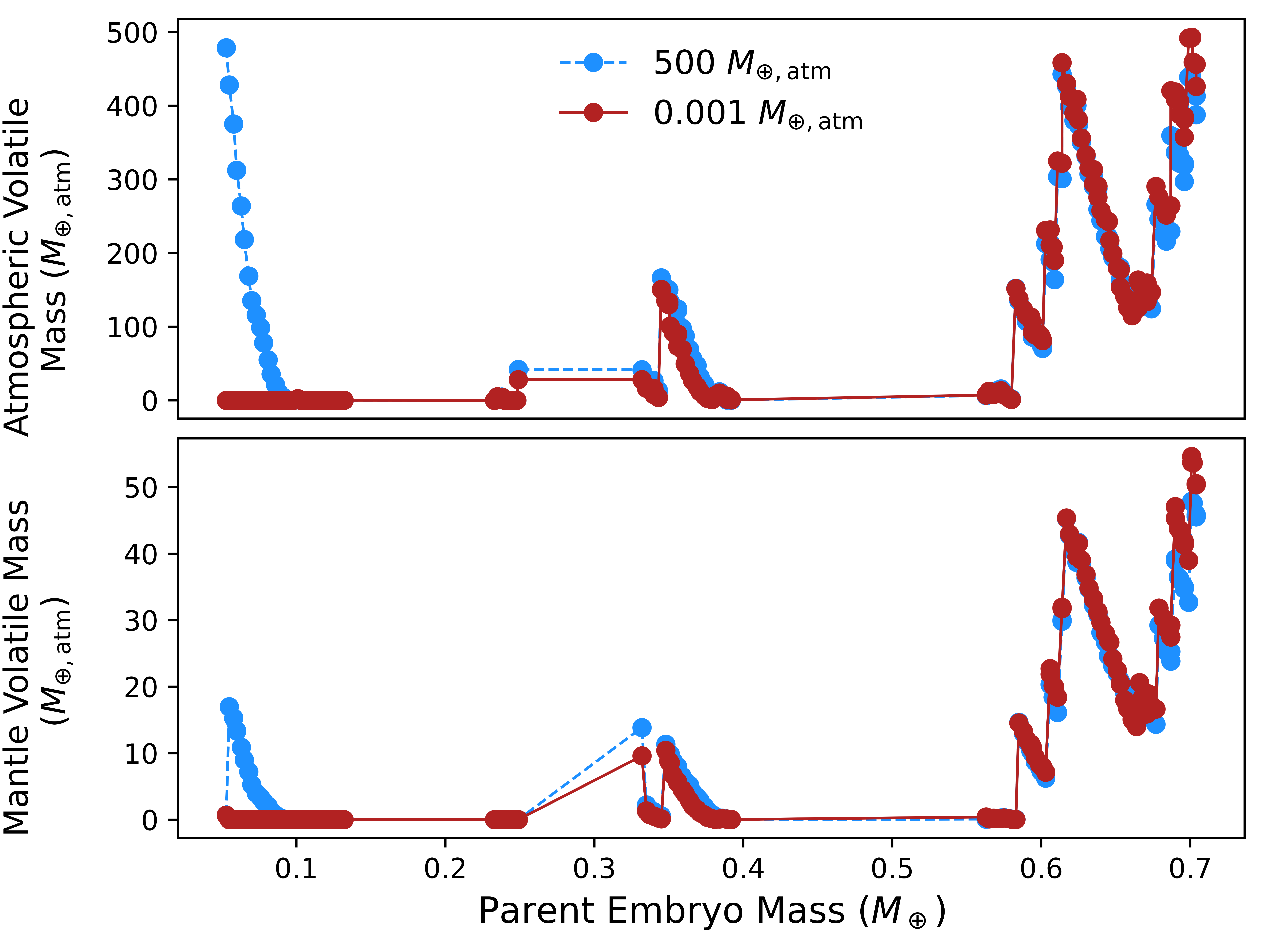}
\caption{\label{fig:plint} Time evolution of atmospheric gas with a growing Earth. Two tracks are shown: one model began with accretion seed of $0.001~M_{\oplus,\rm atm}$ of gas in the mantle and the other $10^4~M_{\oplus,\rm atm}$. Interestingly, they rapidly converge when the planet is only ${\sim} 0.2 M_\oplus$, indicating the existence of one or more equilibrium states that could determine the final properties of the proto-planet.}
\end{center}
\end{figure}  

\subsection{N-body Accretion Simulations: The Grand Tack Paradigm}

We briefly describe the planet migration and formation model used in this study -- The Grand Tack Paradigm, as explored in \citet{WalshEt2011NATURE} and \citet{jacobson+morbi14}. Initially designed to reproduce the architecture of the inner Solar System, the paradigm assumes that Jupiter migrated inwards as Saturn grew. When Saturn approached present-day mass, it started to migrate inwards more quickly, until it captured Jupiter in the 2 : 3 resonance (e.g., \citealt{pierens+14}; finally, the dispersal of the gas disk followed the outward migration of the gas giants.

\citet{jacobson+morbi14} built off the original Walsh et al. model by varying  the total mass in embryos and the ratio of embryo to planetesimal mass (both of these free variables will be important later on this work, as they are the two chief parameters that control dynamical friction). They argued that these values reflect the maturity of oligarchic growth epoch. The reason is that  the  degree  of  evolution  within  the  disk  during oligarchic  growth processes can often imprint itself onto the ratio of embryos mass to  planetesimal mass, or $\Sigma$Me : $\Sigma$Mp.  Overall, their results show that Earth accretion history strongly depends on the assumed total mass and $\Sigma$Me : $\Sigma$Mp.

In this study, we use the outputs of N-body simulation of the Grand Tack scenario described above to place constraints on temporal change of key volatiles on terrestrial Earth-like planets, as well as to explore the effects of accretion history.

\section{Results}   
\label{sec:results}
Using the model described above, we study the volatile growth histories of Earth-like planets by investigating variety of assumptions in initial conditions, impactor composition, impactor population, and accretion history. In particular, we examine the fractionation of CNH elemental species and identify the key processes  driving their resultant compositions.

\subsection{Co-evolution of the Atmosphere and Mantle Reservoirs}

As the model uses N-body dynamical results as inputs, our approach naturally accounts for both planetesimal and giant impacts, as well as their instantaneous, time-averaged, and cumulative effects on volatile accretion. First, we discuss the results for the reference case (Figure~\ref{fig:evo}), which assumes $\Sigma$Me : $\Sigma$Mp = 1, total embryo of $0.5~M_{\rm mars}$, and final planet body of $0.7~M_\oplus$.
Figure~\ref{fig:evo}a shows how the abundances of atmospheric N$_2$, CO$_2$, and H$_2$O change with time reservoir. A few broad observations can be made for the atmospheric abundances (panel a). First, the spikes and dips (i.e., sudden changes in species abundances) are caused by delivery, escape, and degassing. Impact of a large embryo and accretions of volatile-rich planetesimals are primarily responsible for the observed spikes, while successive low mass planetesimals cause the dips in atmospheric mass via efficient impact erosion. Second, volatile gain becomes more pronounced later in time, as seen by the over three times increase in bulk volatile abundance. During the later stages of accretion (when the parent body reaches ${\sim} 0.7~M_\oplus$),  preplanetary materials that originated from beyond 3~AU are being accreted based on the outputs of the N-body accretion simulations. Third, the final atmospheric abundances of N$_2$ and H$_2$O all match present Earth well. The large amounts of calculated H$_2$O is consistent with the formation of a primitive steam atmosphere. Our final surface atmospheric pressure (${\sim} 400$ bar for reference case roughly agree with atmospheric evolution (e.g., \citealt{ZahnleEt1988Icar}) and magma ocean models (e.g., \citealt{Elkins2008EPSL}). The agreement of CO$_2$ abundance between our results and that of modern Earth is poorer, an unsurprising outcome as carbonate crustal block formation is thought to have drawn down massive amounts of atmospheric CO$_2$  \citep{sleep+01b}. 

\begin{figure*}[h] 
\begin{center}
\includegraphics[width=0.6\columnwidth]{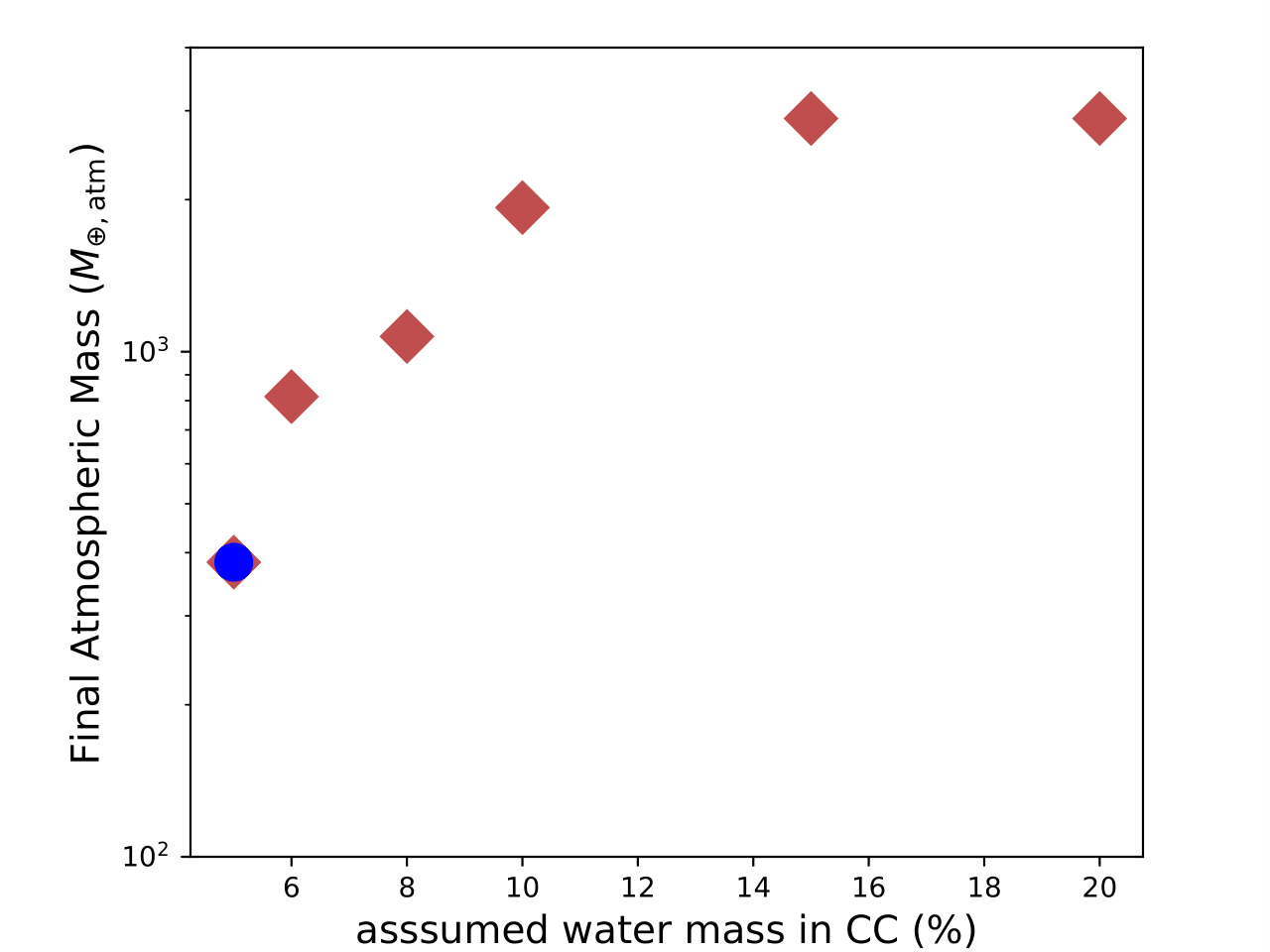}
\includegraphics[width=0.6\columnwidth]{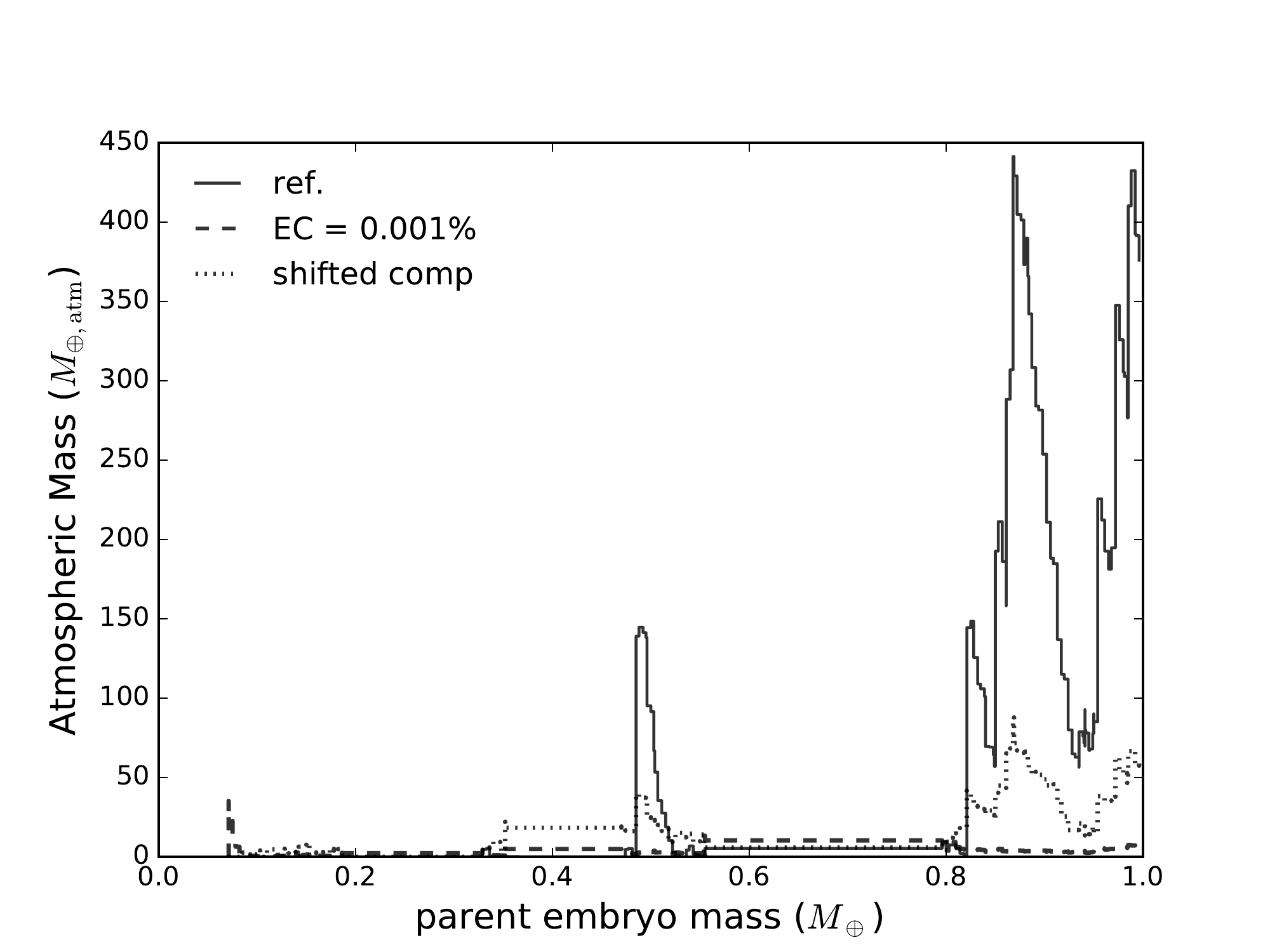}
\caption{\label{fig:plcomp} Sensitivities of atmospheric mass to changes in projectile composition. Here we show the effects of different composition assumptions. In Panel (a), we vary the CC water fraction between 5 and 20\%. Other gases assumed in CCs are fixed at the reference value. Blue circle denotes the reference case, with the most present-Earth-like value. In Panel (b), we show how Earth grows with a flat EC composition and a different composition step function (Equation~1). }
\end{center}
\end{figure*}

The mantle volatile changes similarly to the atmospheric volatiles as a function of time. Yet volatiles in the model mantle are, in almost all cases, less abundant compared to those in the atmosphere (Figure~\ref{fig:evo}b). For example, water content in the mantle is slightly lower that on the atmosphere and that of modern Earth's ocean, but is comparable to previously equilibration models
(see e.g., \citealt{Elkins2008EPSL}). The final carbon abundance is strikingly similar to that of modern Earth. Mantle nitrogen is lower relative to the but within an order of magnitude of measured nitrogen budget \citep{johnson+15}.

Variability in volatile content is directly linked to exchange rates between core-mantle-atmosphere-space. Figure~\ref{fig:esc} shows the total volatile amount eroded (lost), accreted (added), degassed, and ingassed per impact of the reference Earth case. From panel (a), it can be seen that loss events are generally higher and more frequent compared to accretion events. In the rarer events in which volatile accreted (post-impact) dominates, they are usually a few orders of magnitude greater than the volatile lost. These events are caused by larger-sized planetesimals (and in some cases giant impacts), as they are less efficient in removing surface volatiles.  As we will see in Section~3.4, different size distribution of the impactors can cause the ablation-delivery relationship to vary, especially during the later stages of accretion. From panel (b), we see that degassing is almost always greater than ingassing, as the latter only occurs during giant impacts that devolatilize directly onto the magma ocean.  At later times, when the protoplanet reaches ${\sim} 0.6~M_\oplus$, degassing almost completely dominates. The  exception is the time intervals before the parent embryo has reached Mars mass (${\sim} 0.1 M_\oplus$), when ingassing is on average greater than degassing when the mantle was still relatively volatile-poor.

Our model shows that the long-term accretion of volatiles is independent of the initial starting condition (Figure~\ref{fig:plint}). The parent body of the red and blue curves have different initial volatile inventory: $0.001~M_{\oplus,\rm atm}$ and ${\sim}500~M_{\oplus,\rm atm}$ respectively, with everything else held equal. When the parent-body embryo (in this case, an Earth-like planet) is at ${\sim} 0.2 M_\oplus$ however, the two evolution tracks converge and remain qualitatively the same throughout. This behavior implies that there exists certain equilibrium regimes that can established depending on the the properties of the target, its impact history, and the composition of the projectiles. 
We explore the sensitivity of our model to these assumptions in the following sections.

\begin{figure}[t] 
\begin{center}
\includegraphics[width=0.6\columnwidth]{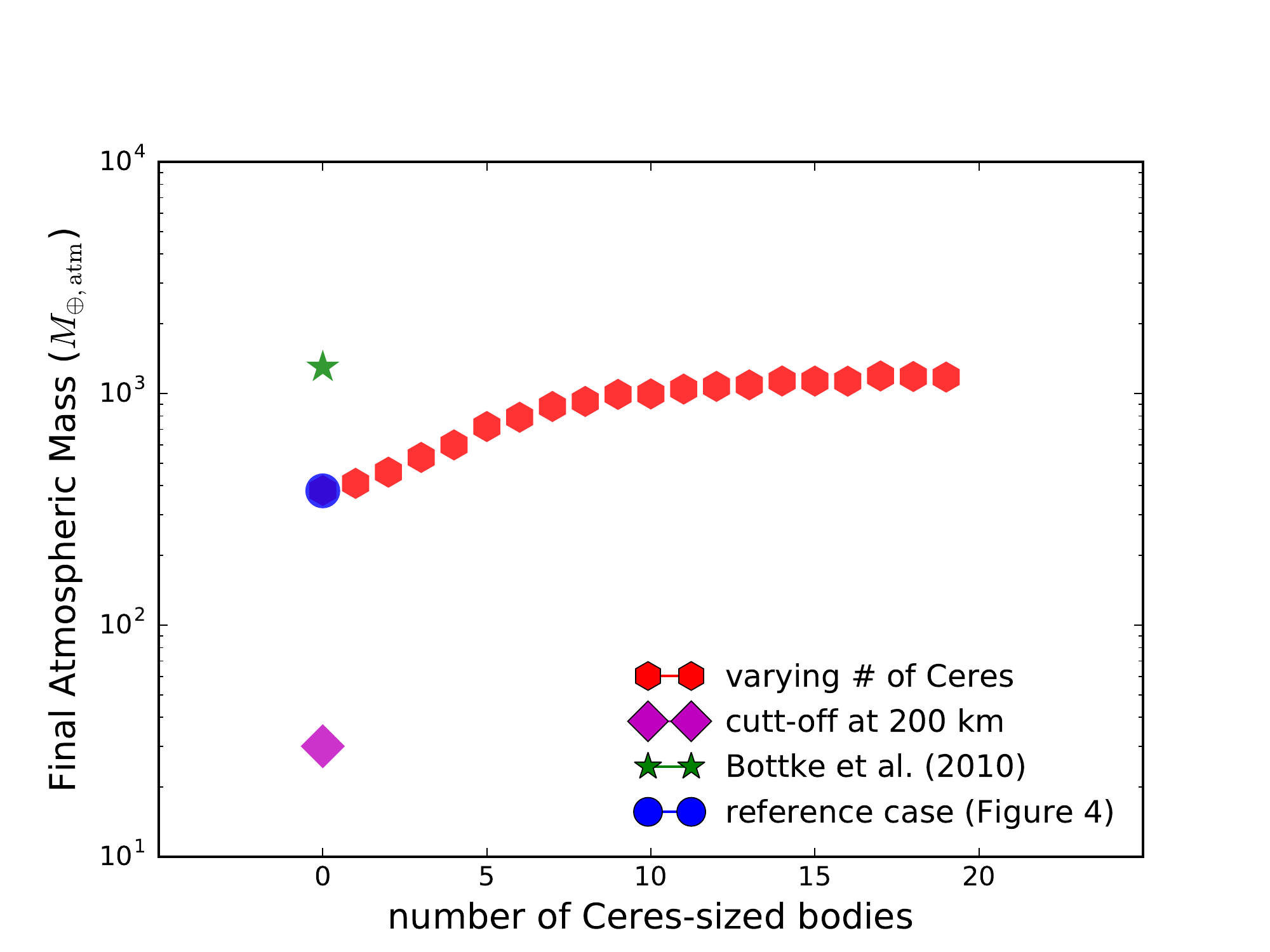}
\caption{\label{fig:pldist} Final atmospheric compositions mass versus planetesimal size, with shape and color indicating the assumed PSD. The labels in the x-axis correspond to the number of Ceres-sized body in the distribution. For hexagons, each assumed successive PSD contains one more Ceres-sized object. The diamond represents a PSD that is capped at 200 km. The star represents a PSD characterized by massive projectiles (between 1000 and 3000 km in diameter) as proposed by \citet{BottkeEt2010Science}. Blue circle denotes the reference case, which happens to most accurately match the properties of present Earth (Figure~\ref{fig:evo}) }
\end{center}
\end{figure}

\subsection{Importance of Planetesimal Compositions}
 
To explore a reasonable spectrum of chondrite chemical compositions, we perform additional simulations involving changes in: 1) volatile-fraction of enstatite and carbonaceous chondrites, 2) location of step function cut-offs.   

Figure~\ref{fig:plcomp}a shows results where CC water fraction is varied between 5 and 20\%. The reference 5\% value roughly reproduces an atmospheric pressure value predicted by other models, as noted earlier (e.g., \citealt{ZahnleEt1988Icar}). Unsurprisingly, the higher water content assumed, the more massive the final steam atmosphere becomes. As the CC fraction is raised to ${\sim}20\%$, the final atmospheric masses markedly flattens out at about $4000~M_{\oplus, atm}$. 

In Figure~\ref{fig:plcomp}b, three runs along with the reference case (solid line) are shown. First, the step function assumption in Equation~1 is modified to the following:

\begin{equation}
\chi_{\rm pl} =\begin{cases} {\rm E{\rm -}type } & {\rm for} \hspace{2 mm}  a < {\bf 1.5} \\
                     {\rm S{\rm -}type }  &  {\rm for} \hspace{2 mm}   {\bf 1.5} < a < {\bf 3.5} \\
                     {\rm C{\rm -}type } & {\rm for} \hspace{2 mm}   a \geq {\bf 3.5}
       \end{cases}
\end{equation}

\noindent where $a$ is the heliocentric distance and the bolded numbers denotes the changes made from Supplemental Information Equation~1. With these shifts, we have effectively expanded the heliocentric width for ordinary chondrites, while reducing the width for enstatite and carbonaceous chondrites. The results for this simulation is indicated by the dotted curve.
The other curve (dashed) represents the assumption of a pure EC composition ``flat" distribution (i.e., homogeneous accretion). Both runs resulted in lower atmospheric pressures than the reference case. This outcome illustrates the key role of CCs (planetesimals from greater orbital radii) in determining volatile abundances. Although some amount of enstatite chondrite is replaced with the more volatile-rich S-type, the amount of CCs accreted when the parent embryo is ${\sim} 0.8 M_\oplus$ is significantly reduced. It is interesting to note that before $M_{\rm parent} \sim 0.5~M_{\oplus, \rm atm}$, the reference case is never the highest in atmospheric mass despite having the highest final value. Several explanations may be behind this behavior. First is the dependency of mass-loss efficiency on mean molecular weight $\mu$ of the atmosphere through dependency on the scale height ($r_{\rm cap} \propto  \left(H_{\rm tar} R_{\rm tar}\right)^{1/2} \propto 1/\mu$).   Second, before the parent body reaches ${\sim} 0.5~M_{\oplus, \rm atm}$, volatile-rich body collisions are infrequent, and  any small increase in volatiles is quickly eroded, thereby suppressing net accumulation. Note that  while the outer solar system planetesimals get scattered by Jupiter's inward migration before the parent body reaches ${\sim} 0.5 M_\oplus$, they do not necessarily get accreted. This is because these scattered objects are on much excited orbits, leading to high relative velocities and eccentricities; the resultant low gravitational cross section enhancement makes direct accretion by the growing embryo more difficult.

\begin{figure*}[t] 
\begin{center}
\includegraphics[width=0.8\columnwidth]{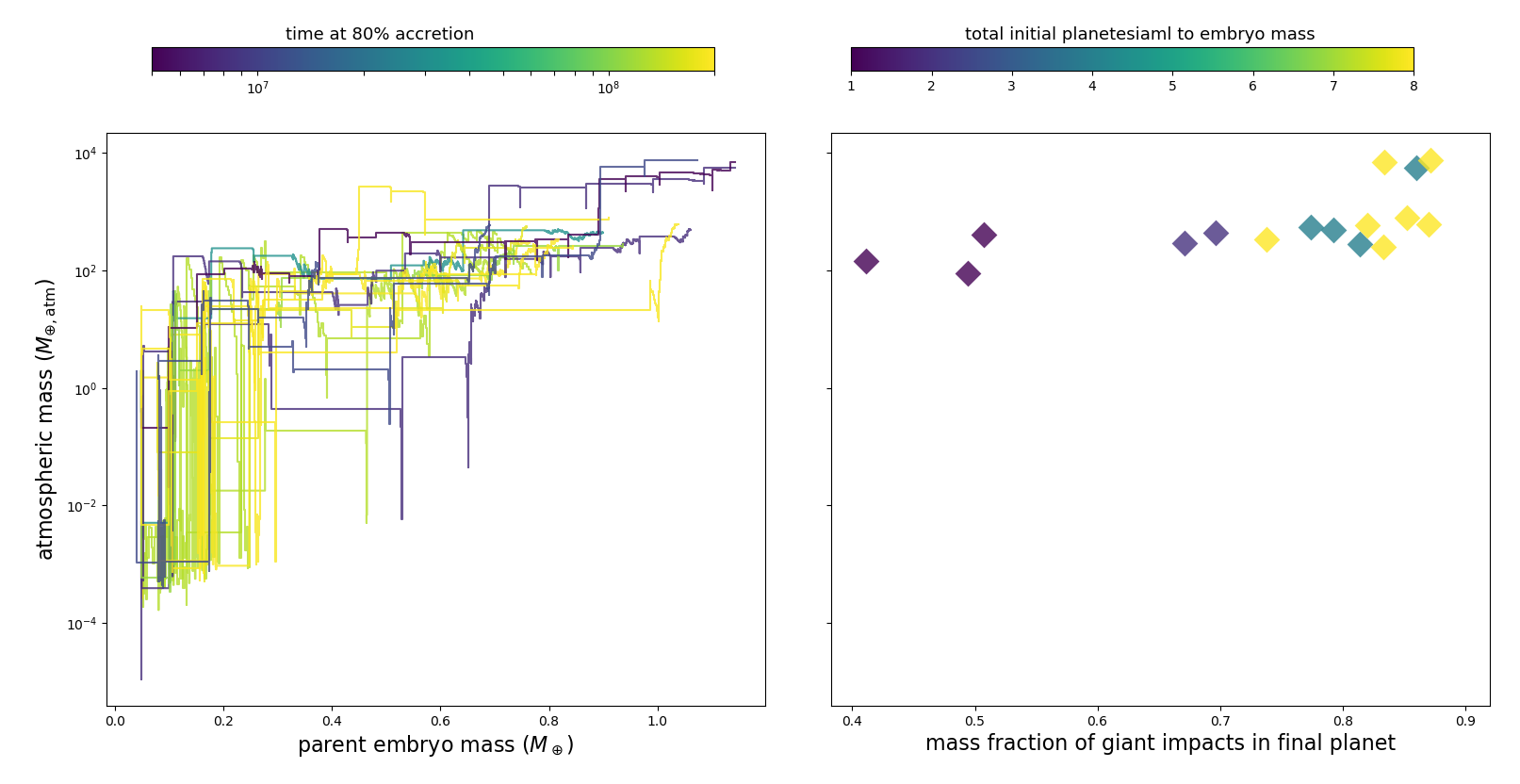}
\caption{\label{fig:nbody} Atmospheric growth from a suite of N-body simulations to demonstrate the effects of accretion history. Panel (a) shows atmospheric growth histories using all 16 N-body simulations. In panel (b), we show the dependence of final atmospheric mass (when the protoplanet reaches $1~M_\oplus$) on the giant impact mass fraction in the final planet} and the total initial embryo mass in the simulation. Blue circle denotes the reference case, which happens to most accurately match the properties of present Earth (Figure~\ref{fig:evo}).
\end{center}
\end{figure*}

\subsection{Influence of Planetesimal Distribution Assumptions}
In the reference simulation cases, we have drawn from the asteroid-belt population to determine the sizes of each planetesimal. Here we test the effects of varying planetesimal size distribution (PSD) assumptions. We perform two modifications to our current planetesimal population--a) including additional Ceres-sized objects,  the volatile abundance of which will follow the initial heliocentric distance and not the measured composition of Ceres and b) capping the upper limit of the size distribution at 200 km in radius, c) adding more giant bodies between 500 and 1500 km in radius, mimicking the size distribution used by \citet{BottkeEt2010Science}.

The results of these changes to the PSD are shown in Figure~\ref{fig:pldist}. From the red circles, we see that the addition of Ceres-sized (${\sim}500$ km) bodies results in increased atmospheric mass. This is due to the more dominant contribution of small planetesimals to the total mass-loss history (Figure~\ref{fig:esc}). With more large Ceres-sized bodies in the distribution, fewer planetesimals with $r_{\rm pl} \sim r_{\rm min} \sim \left(3 \rho_{\rm surf}/\rho_{\rm pl}\right)^{1/3} H_{\rm tar}$ make up each super-planetesimal. In the case of the \citet{BottkeEt2010Science} PSD, we find that the result is similar to adding several (between 15-20) Ceres-sized bodies to the asteroid population, namely a substantial increase in the final volatile mass.  This result indicates that a distribution with a longer tail than the nominal asteroid population is unrealistic as the assumption produces extreme ($> 1000$ bar) atmospheric pressures.

Conversely, without any planetesimals greater than 200 km, the final atmospheric mass is dictated by sub-10 km projectile impact escape. This leads to a suppression of atmospheric growth (to ${\sim} 30$ bar).

\begin{figure}[h] 
\begin{center}
\includegraphics[width=.7\columnwidth]{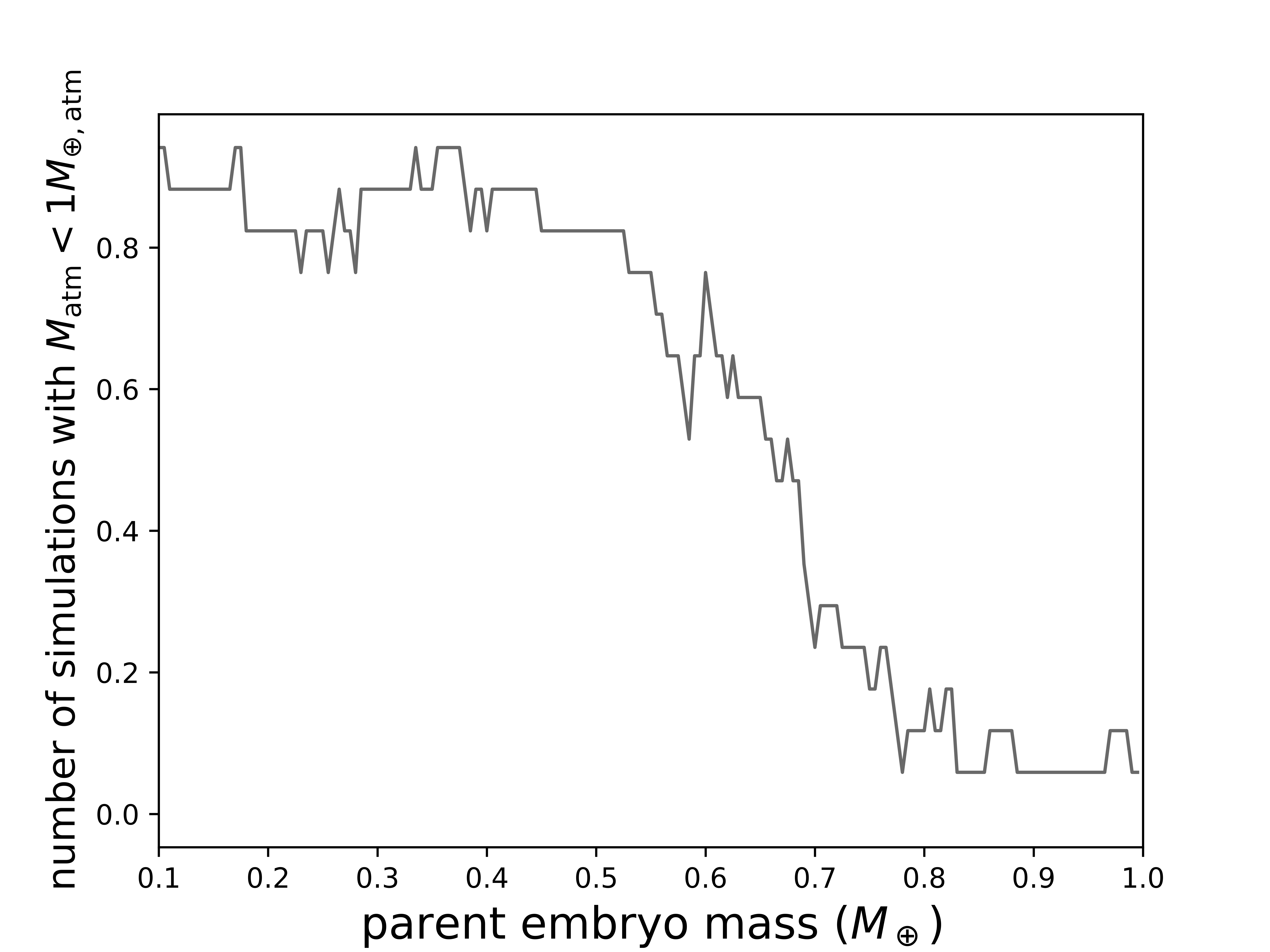}
\caption{\label{fig:frac}  Fraction of simulations with atmospheric masses less than $1~M_{\oplus, \rm atm}$ as a function of the parent body mass. A total of 16 N-body simulations was performed, each with different ratios of total planetesimal mass to embryo mass and individual embryo mass. We find that the number of simulations with $< 1~M_{\oplus, \rm atm}$ decreases with time, suggesting enhanced efficiency of volatiles gain in the later stages of accretion. }
\end{center}
\end{figure}  

\subsection{Trends in Multiple N-body Simulations: Effects of Accretion History}
We test the effects of accretion history on our results using outputs from a suite of N-body accretion simulations. In each N-body simulation, the only changes are the total planetesimal to embryo ratios and the individual embryo mass. 

The outcomes of these 16 N-body simulations are displayed in Figure~\ref{fig:nbody}.  In Panel (a), we show the time evolution of our model using these different N-body simulation outputs. Due to the different initial N-body simulation conditions, each growth curve follows a distinct track. However, embryos that reach 80\% of its final mass in a shorter period of time have typically more massive atmospheres during accretion. This is seen by the fact that none of the simulations with accretion time ${\sim} 10$ Myr drop below $30~M_{\oplus, \rm atm}$ after $M_p > 0.35~M_\oplus$.  Intuitively, we can understand this as an interplay between dynamical accretion and delivery efficiency of the impactors. Systems that grow quickly are those that have more dynamical friction, increased accretion rate, mass fraction of giant impacts (Figure~\ref{fig:nbody}), and final atmospheric mass. For embryos that quickly accumulate to near-Earth-mass, the ability for impact losses to erode these atmospheres is substantially lessened. 

In Panel (b), only the final atmospheric masses are shown (as a function of total mass of CC in final assemblage) for each simulation. Some dependence on the total CC accreted can be seen. This is because CCs are very abundant in volatiles, so the more that is contained within a planet simulation the more volatile-rich it will become, assuming the same amount of volatile-loss. As CC accretion fractions $> 5\%$, final atmospheric masses begin to plateau at close to 1000 $M_{\oplus, \rm atm}$. There is also a slight dependence on the initial planetesimal to embryo mass ratio, as seen by the clustering of colors for simulations with higher planetesimal to embryo masses.

%The total water content predicted by our calculations is between . The majority of these final values are consistent with estimates of Earth's global water content. For instance, 

Figure~\ref{fig:nbody} demonstrates that impact history matters in determining the final atmospheric mass and composition, but are there any identifiable trends across these results? We now turn to evaluating atmospheric growth properties of the ensemble. To do this, we record the atmospheric mass for each simulation at each timestep and determine if the value is lower than 1 $M_{\oplus, \rm atm}$. As shown in Figure~\ref{fig:frac}, fluctuations in the number of simulations with light atmospheres occur due to the competition between impact erosion and delivery. The general trend however is that the number of simulations with low atmospheric masses decreases as the accretion proceeds, suggesting that our protoplanets experience a more pronounced net volatile gain at the later stages of accretion. We find that the majority of our results are consistent with early predictions by \citet{ZahnleEt1988Icar}, namely our mean surface atmospheric pressure typically fall between ${\sim}200 - 400$ bar.

\begin{figure}[h] 
\begin{center}
\includegraphics[width=.7\columnwidth]{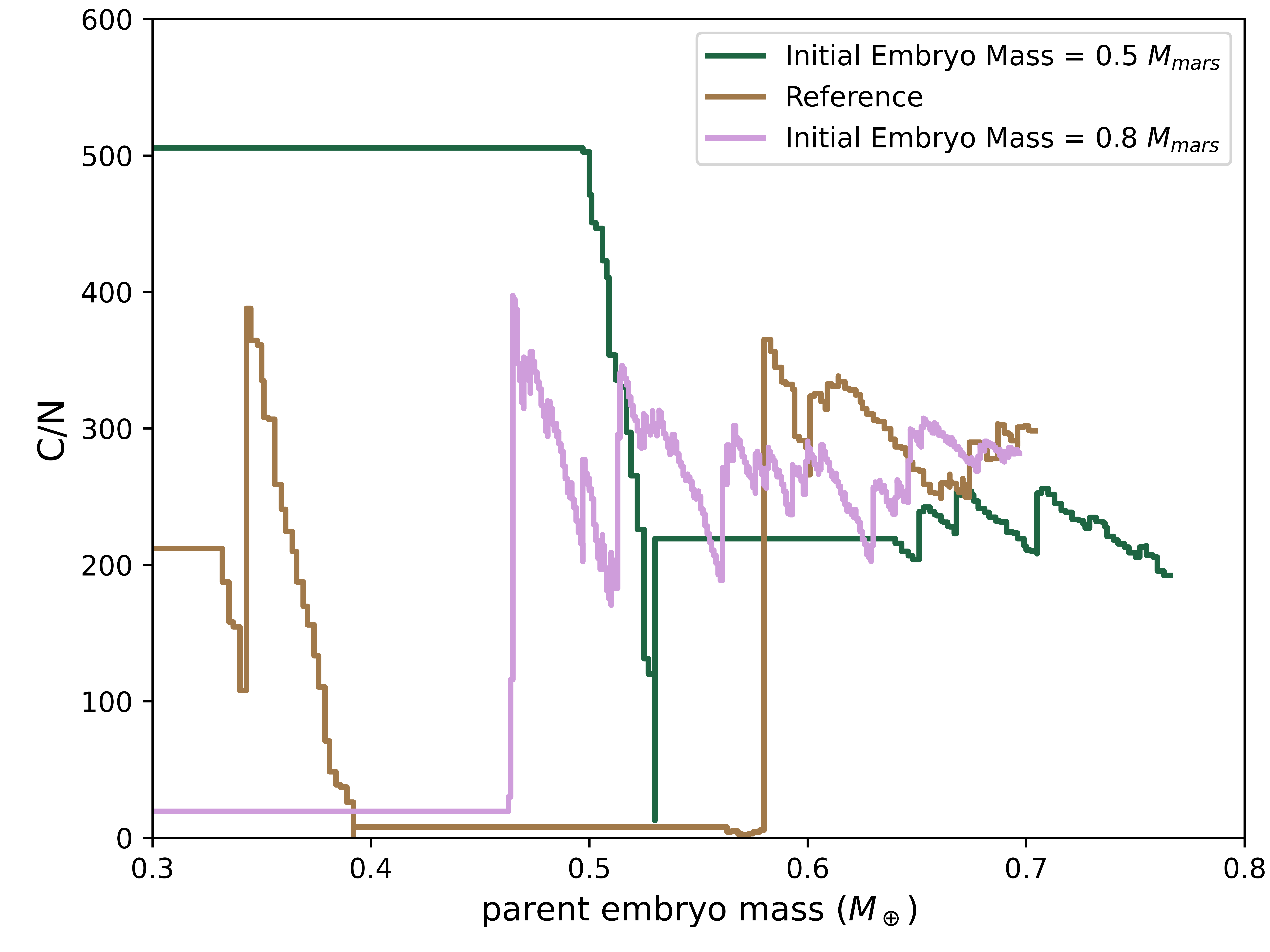}
\caption{\label{fig:cnr}  Sample temporal evolution of C to N ratio of the parent body, showing that the superchondritic C/N arises from the preferential atmospheric loss and mantle degassing of nitrogen over carbon.  }
\end{center}
\end{figure}  

\subsection{C to N Ratios as a Natural Consequence of Accretion Processes}

%make stronger, stream line this
We show here that superchondritic BSE C/N is an emergent property of planet formation. This is seen by the fact that all our model runs ended with  high final C to N ratios relative to the chondrites, even with a spread in the source chondritic composition (with C/N varied from 10 to 25; Supplementary Table~2). From the temporal evolution of C/N in the three  simulations shown in Figure~\ref{fig:cnr}, we see that the rise in C/N occurs almost immediately after the first impacts and only briefly return to their initial chondritic values (before the embryos have grown to $0.6~M_\oplus$). These results suggest that for typical planetary embryos, superchondritic C/N can be achieved with a set of reasonable boundary conditions. 

\begin{figure*}[t] 
\begin{center}
\includegraphics[width=0.8\columnwidth]{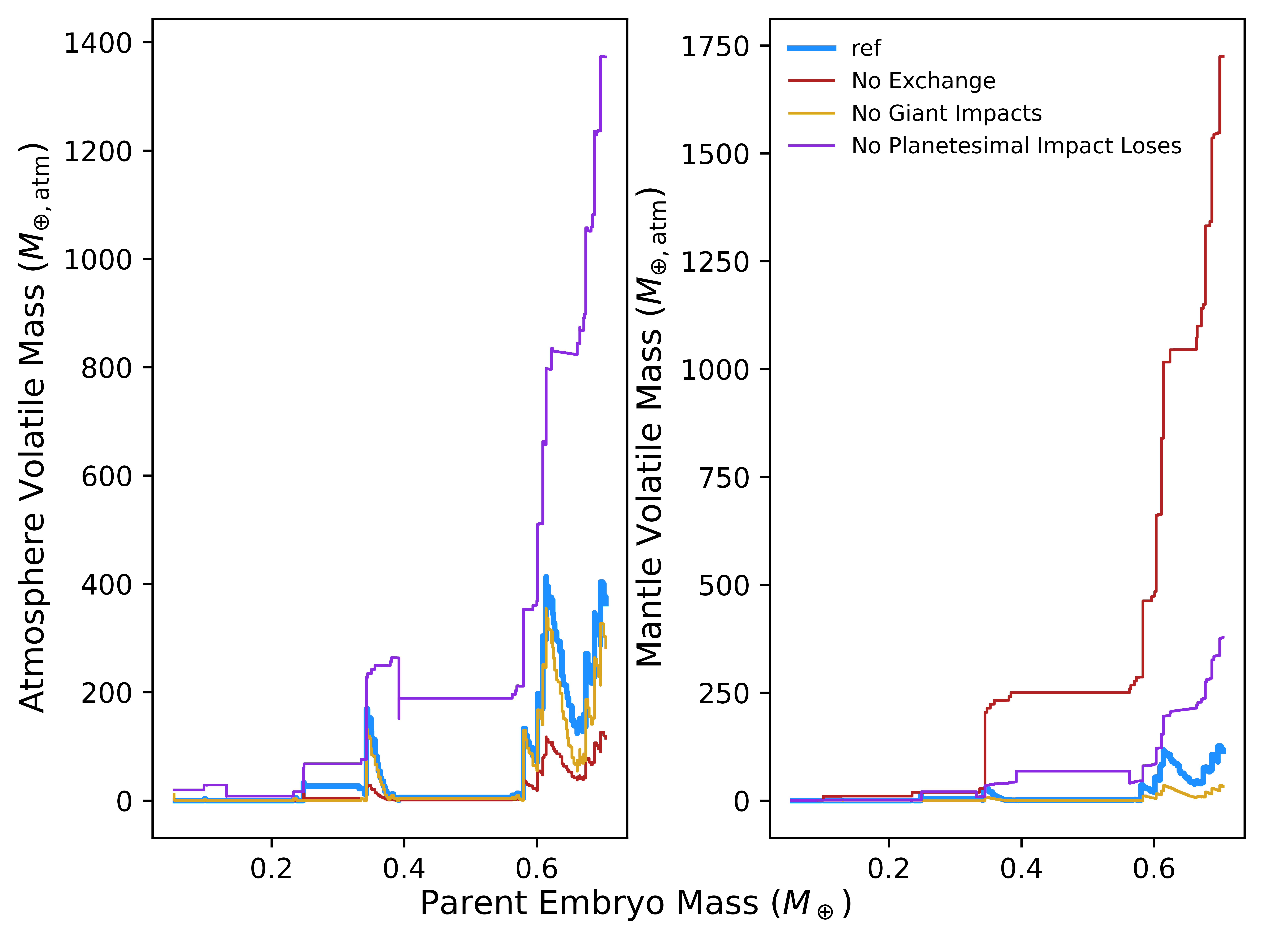}
\caption{\label{fig:impacts} Time evolution of atmosphere and mantle volatile inventory as an Earth-like planet grows, normalized by their respective present-day atmosphere (left) and mantle (right) values. Each panels include results a) without the effects of planetesimal impact erosion (red), b) without the effects of giant impacts (gold), and c) with both erosion and giant impacts (blue). }
\end{center}
\end{figure*}

To examine the role of different impact events in affecting the final C, N, H abundances, we test additional model scenarios without the inclusion of giant embryos and impact erosion by planetesimals. We find that the neglect of giant impacts still lead to high C/N, yet ignoring planetesimal impacts, in nearly all cases, lead to similar or lower C/N ratios as the impactors themselves, implying that both loses to space and the subsequent mantle-atmosphere exchange are necessary to produce the superchondritic C/N. Furthermore, in the scenario without the inclusion of impact loses, the final atmospheric mass is much elevated (in excess of ${\sim}1500 M_{\oplus,atm}$; Figure~\ref{fig:impacts}). As discussed in Section~3.2, such a large primitive atmosphere is not realistic as it would be challenging for hydrodynamic and impact erosion to reduce the atmosphere mass to the present-day value. In contrast, ignoring giant impact slightly reduces the final atmospheric mass, indicating that giant impacts play a volatile-delivery rather than volatile-depletion role. Note however, that the presence or non-presence of an ocean during a giant impact is not accounted for in the current model. In the presence of an ocean, they can be more efficient than what is calculated here.

\begin{figure*}[t] 
\begin{center}
\includegraphics[width=.6\columnwidth]{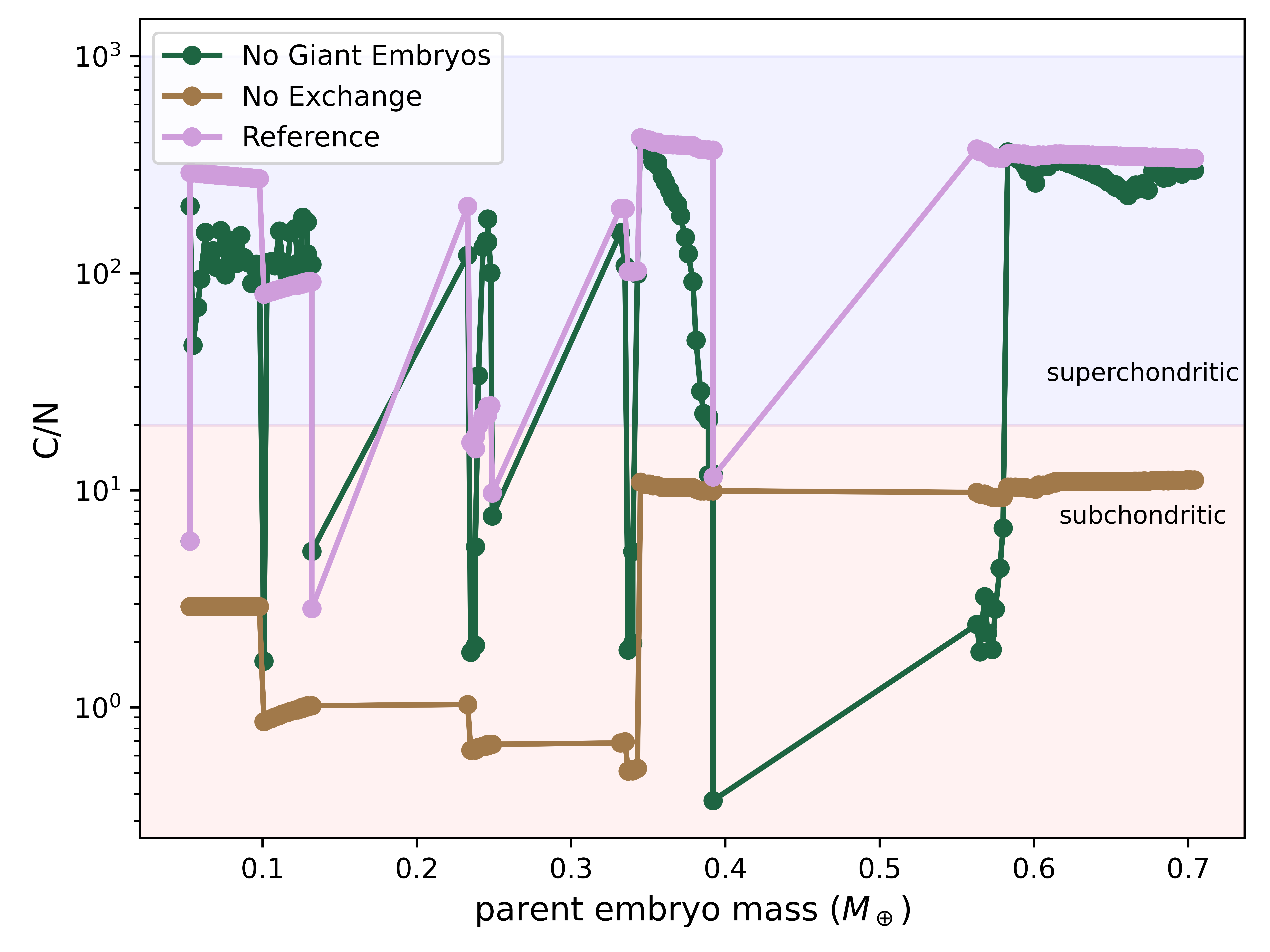}
\caption{\label{fig:cnr2} Time evolution of C to N ratio as an Earth-like planet grows. Different assumptions are represented by their colors. We show that ignoring either giant impacts or mantle-atmosphere exchange result in C to N ratios that are incompatible with geochemical records.    }
\end{center}
\end{figure*}

Our results indicate that the synergistic effects of impact erosion and degassing lead to superchondritic BSE C/N ratios. To more quantitatively illustrate the magnitude of each potentially competing process, Figure~\ref{fig:cnr2} shows that the continuous  mantle degassing of nitrogen over carbon during the later stages of accretion results in high C/N. N is less solubility in the magma ocean compared to C. Hence, mantle nitrogen is more rapidly drawn out (relative to carbon) as evidenced by the enhanced degassing rates. On the other hand, the higher escape rates of CO$_2$ is somewhat misleading due to the higher delivery rate of carbon via  CI chondrites, thus creating atmospheres richer in CO$_2$.

%note on C/H
%Finally, we note that our results produce subchondritic C/H, consistent with geochemical data (Figure~\ref{fig:cnr2}). 

%\begin{figure}[t] 
%\begin{center}
%\includegraphics[width=1.\columnwidth]{Figure_11a.png}
%\includegraphics[width=1.\columnwidth]{Figure_11b.png}
%\caption{\label{fig:rate} Net gained (delivered -  lossed) and net outgassed (degassed - ingassed) N and C budgets during %each impact event.  }
%\end{center}
%\end{figure}  

\section{Discussion}
\label{sec:discussion}

\subsection{The Elemental Ratio Conundrum}

On Earth, both extremely and moderately volatiles are depleted relative to CI chondrites and other planetary precursors, but the isotopic signatures of the BSE point to an inner solar system asteroidal origin \citep{halliday+13,alexander+12}. 
Previous work show that admitting a small amount of cometary materials allows the noble gas proportions to match that of the Earth \citep{halliday+13}. Others, using data from 67P/Churyumov-Gerasimenko, argue that the bulk of present day Earth only has minimal cometary contribution \citep{bekaert+20}, but the added materials does not resolve the severe fractionation of moderately volatile species (C, N, H, S) on Earth. In fact, no known Earth reservoir signature perfectly associate with any known chondrites, primitive nor processed ( Though see \citet{Piani+2020}, where they suggested that the majority or all of Earth's water and nitrogen may have come from enstatite chondrites).

To resolve this apparent conundrum in elemental composition between Earth and the putative planetary precursors,  recent work found  that differentiation of preplanetary bodies could transfer carbon to cores during the small solid body processes \citep{hirschmann+21}, a channel that could explain both the chondritic C/S and subchondritic C/H. The origin for the observed BSE C/N is likely to be less straightforward, however, as the C/N values of the interstellar medium, the early nebula, comets, and various types of chondrites differ substantially (i.e., from ${\sim} 5$ to ${\sim} 200$; \citealt{bergin2015pnas}). 
Interestingly, ordinary chondrites have a diversity of C/N values, a subset of which approach the BSE C/N.  Once again, this does not mean that ordinary chondrites are the primary constituent of Earth, as their isotopic signatures do not agree. But it does suggest that one or a combination of metamorphic alteration processes are at work, resulting in the observed range of inherited C/Ns. 
%Measured C/N in the mantle reservoirs vary widely (whereas the surface and crustal C/N are well constrained by geologic record (to ${\sim} 21$ and ${\sim} 60$ respectively)), and is likely due to the variety of metamorphic processes and the need to accurately correct for atmospheric and fractionation contributions.

Previous work has shown that core formation is more likely to reduce rather than raise the BSE C/N due to the strong siderophile nature of carbon \citep{hirschmann2016am,dalou+17}, and the similar mantle C and N behaviors \citep{grewal+20} would promote subchondritic C/N.  \citet{hirschmann2016am} insofar argued  that the draw down of carbon to the core is exceedingly efficient to such an extent that even significant removal of N cannot counteract the reduction of C/N. 
In another study, \citet{grewal+19} tested different volatile fractions for their giant impactor and found that superchondritic C/N could arise during the merging of a differentiated embryo with an E-type-chondritic composition\footnote{Note that they found that the atmosphere could also become N-rich if an S-rich core can be achieved, making N$_2$ less soluble in the MO.}. 
However, the majority of these studies assumed static single-stage single-reservoir magma oceans.  While we adopt a similar mantle plus an overlying atmosphere interior model, the inclusion of stochastic accretion allows the investigation of  chaotic processes and the competition between volatile delivery and escape. One key feature of our results is that the predicted final C and N inventories do not hinge on the assumed composition of the preplanetary materials; superchondritic mantle C/N is found regardless of the details of the dynamical history and volatile fraction of the planetesimals. Hence both impact induced degassing and accretion of a late veneer (i.e., differentiated materials) may play roles in elevated the BSE C/N \citep{bergin2015pnas}.

Our study highlight the key role of time-dependent impact losses in driving the elemental composition of nascent worlds. Our results favor a scenario more consistent with the those posited by \citet{bergin2015pnas} and \citet{tucker+14}. In particularly however, \citet{bergin2015pnas} argued that during very reduced conditions, the atmosphere can be CO$_2$-dominated and the nitrogen can be sequestrated as metals, thereby creating high BSE C/N. However, this suggestion neglected the continual effects of impact erosion. During the later stages of accretion, our work shows that rapid loss of nitrogen in oxidizing (rather than reducing) conditions should eventually lead to an elevated mantle C/N.

%Compare C/N in marty, surface (atm), and mantle res, compare with bergin, note similar pattern between mantle C/N and surface C/N trends between our results and marty's.    comment on C/H and C/S

Because elemental fractionate in accordance with their solubility coefficients, and the solubilities depend on the melt composition (and therefore the bulk planetary composition), the calculated C/N will vary with the planet in question. A future study will study the elemental compositions of Venus and Mars and their extrasolar counterparts, where life-relevant ingredients may be commonplace. Indeed, detection of circumstellar materials (e.g., ``exo-Kuiper-belts") around white dwarf has shown the prevalence of key element such as C, N, O, S \citep{xu+17}. An important caveat when considering other planets is that the mantle-atmosphere elemental partitioning behaviors could also be highly $P$- and $T$-dependant \citep{Roskosz+2013,dalou+17}, pointing to the need for further laboratory constraints.

\subsection{Volatile Sinks: What Dominates?}

%core sequesttion vs atmosphere escape

During planetary accretion, volatiles transfer from the original nebular component to the various reservoirs on a planetary body. Much of these  are then lost during further planetary processing. Core formation, MO crystallization, and  parent body metamorphism and partial differentiation, for example, incur significant losses for  moderate volatiles \citep{dalou+17}, noble gases \citep{marty+20}, and even highly siderophile elements such as carbon  \citep{hirschmann+21}. Here, we find  that the dominant loss process is impact erosion, particularly in later stages of accretion during high influxes of volatile-rich planetesimals.  %Volatile exchange between the atmosphere and the underlying magma ocean.

% giant impacts vs planetesimals, what's more important
Giant impacts are thought to substantially influence the global volatile inventory of forming worlds \citep{kegerreis+21,Genda+Abe2003Icar,Genda+Abe2005NATURE,grewal+19}. %giant impact dynamics and delivery, cite
This study suggests that giant impacts do not play as of an important role as previously suggested; excluding the effects of giant impacts only led to a ${\sim}20\%$ change in the final atmospheric mass (Figure~\ref{fig:impacts}).  By randomly sampling planetesimal sizes from modified versions of the asteroid-belt distribution, we find that the occasional accretion of large (km) planetesimals acts as the key controls of the final volatile abundance. More specifically, planetesimals masses drawing from an asteroid-belt-like distribution  best fits the outcomes in this study. In turn, our results strongly disfavors the PSDs with many large sized bodies, e.g., one posited by \citet{BottkeEt2010Science}. This is because the increased stochastic arrival of large-sizes bodies increases the total mass of the atmosphere by more than two orders of magnitude. This result is consistent with the earlier predictions of \citet{deniem+12}, where they found that large and slow impactors are particularly important for the final atmospheric mass.  To reiterate however, these conclusions will likely depend on the presence of an ocean, which our study does not explore. 

Hydrodynamic atmospheric escape approximated with the energy-limited formalism does not substantially contribute to volatile loss. However,  recent work has found that the energy-limited approximation have been over/under-estimate the mass loss rate by over two orders of magnitude \citep{krenn+21}, and thus explicit simulations of (magneto-)hydrodynamics should ideally be incorporated. Furthermore, energetic astrophysical event such as stellar flares and coronal mass ejections may enhance stratosphere/mesospheric atmospheric moisture \citep{chen+19,chen+21} and potentially lead to the long-term dehydration of the upper atmosphere in young systems \citep{dong+17,green+21}. Hence solar (stellar) activity (as well as those of other astrophysical events; \citealt{ambrifi+22}) is implied to dramatically impact on evolved (exo-)planetary atmospheres, but their effects on early volatile growth warrants further numerical quantification.

\subsection{Model Limitations \& Future Problems}

As a first study to integrate the multitude of processes into volatile accretion, here we address a list of simplifications and assumptions made that readers should keep in mind. We will use these limits as a guide for future work.

First, we did not account for changes in magma ocean depth and the degree of core mantle equilibration associated with each impact. Compared to small planetesimal accretion, large impacts equilibrate at higher pressures and are able to influence a planet's deep interior. Second, for the sake of simplicity, we did not calculate how magma ocean depth varies as a function of impact characteristics. To better delineate the location and extent of metal-silicate equilibration, we will use \citet{Nakajima+21}’s scaling law to calculate heat distribution and melt fraction for each impact.  Third, we have treated core formation, magma ocean degassing, and atmospheric escape separately, in a sense that they happen sequentially one followed by another. This is clearly non-physical as these processes are expected to occur simultaneously.

Because our hydrogen species resides in H$_2$O and carbon species in CO$_2$, one inevitable  is the assumption of a constant oxidizing condition. Temporal variations in redox, which we plan to implement in a follow-up work, will affect volatile speciation and hence its degassing history. In addition, we disregarded the presence of an ocean present during a giant impact; the presence of an ocean would amplify atmospheric loss rates by a giant impact (e.g., \citealt{Genda+Abe2005NATURE}). However, the amount of ocean present depends on the time interval between impacts versus the time to condense a steam atmosphere.  

Finally, the current core model only contains carbon. N will unlikely be incorporated into the core with the exception of extreme high temperature conditions. H will only be sequestered into the core if the mantle is under strong reducing conditions, which might be relevant for early times. Given that we assume relatively mild oxidizing conditions (more compatible with later stages; \citealt{rubie+15}), not much H will be found in the core. These different assumptions for core composition and evolution will be examined in a later study.

%%future problems below
This study aims at constraining the time dependent growth and loss of major gaseous species that constitutes modern Earth's atmosphere (N$_2$, H$_2$O, CO$_2$), but did not include extremely volatile noble gases (Ne, Ar, Kr, Xe). Noble gases are advantageous due to their high ionization energies, reflecting their low chemical reactivity. They also tend to almost always remain in the gaseous phase, making them the most important atmophiles. These properties makes noble gases exceptional as tracers for the origin and history of fluids (see e.g., \citealt{Pepin1997Icar,PorcelliEt2001EPSL,tucker+14}).

As an extension to our code, we will include noble gases such as He, Ne, Kr, and Xe, which will allow us to investigate, e.g., the Ne/He system. This system is remarkable because whereas many noble gases Ar, Kr, Xe are vulnerable to recycling in planetary interiors, He and Ne are recycled only in minor quantities (e.g., \citealt{HollandEt2006NATURE}). Mid-ocean ridge basalts (MORB), ocean island basalts (OIB), have significantly different $^3$He/$^{22}$Ne compared to the nebular component. %Several possible explanations have been suggested. Plate tectonics, for one, could recycle and fractionate these gases in the interior. \citet{tucker+14} on the other hand explained this discrepancy by episodes of  MO degassing driven by giant impacts, leaving the primitive MO elementally fractionated. As pointed by \citet{SchlichtingEt2015Icar} however,  impacts large enough to completely strip the atmosphere most certainly will generate large-scale melting. The lack of geochemical evidence for global magma oceans is problematic for the giant impact argument. %For this reason, one wonders whether if other mechanisms could be responsible--a straightforward one could be that not accounting for these mass-loss episodes could lead to severely overestimate net volatile-accretion, which would disfavor more degassing. We will investigate this possibility with our own N-body + atmospheric growth model.
The $^{30}$Ar/N of the BSE and CV (and CO) chondrites are also enhanced compared to CI-CM chondrites. Another possibility for the observed depletion stem from the variable thermal threshold of the gas-bearing hosts \citep{marty+20}, as noble gas-bearing (and often more refractory) materials are affected at higher temperature than nitrogen-bearing ones. When in contact with high temperatures then, nitrogen can be more easily liberated and subsequently lost from the less refractory organics.  We plan to explore these potential scenarios by quantifying the volatile retention rates of different host phases.

To allow our present model to calculate atmosphere and mantle oxidation chemistry, two major modifications are needed. First, we will need to add more ``boxes" to the mantle reservoir (e.g., \citealt{PorcelliEt1995GCA}). While a good approximation, typical mantle-core and mantle-atmosphere equilibration do not involve the whole the MO, with the equilibrium pressure loosely dependent on the impactor mass \citep{armstrong+19}.
Second, physical mixing and transport can fractionate different noble gas isotopes. To account for noble gas isotopes, we will need to implement absorptive, hydrodynamic, and Rayleigh fractionation models
(e.g., \citealt{Pepin1991Icar}). Apart from C, N, and H, we will include noble gases He, Ne, Ar, Kr, Xe and stable isotopes (e.g., $^3$He, $^{22}$Ne, $^{38}$Ar, $^{40}$Ar, $^{129}$Xe).

Oxygen fugacity ($f_{\rm o2}$) is a crucial parameter that regulates the partitioning of siderophile, lithophile, and atmophiles. In this work, we used fixed (but species-dependent) Henrian solubility coefficients to calculate the distribution of gases in the mantle versus the atmosphere. As Earth-like planets originate in highly reduced conditions and begin to oxidize when H$_2$O-bearing materials are being accreted \citep{rubie+15}, future work will need to study extent to which these materials could raise the oxidation state of proto-Earth. In addition to oxidation via H$_2$O dissociation and H$_2$ release, segregation of metallic iron and Fe$^{3+}$ to the core can strongly oxidize the upper mantle \citep{armstrong+19}, which could facilitate the outgassing of oxidized species. The effects on mantle and atmospheric oxidation, particularly during large impacts, will need to be evaluated with self-consistent calculations of $f_{\rm o2}$ evolution and its effect on volatile abundance and equilibration. This effort is especially important as similar intrinsic oxidation states to Earth, Mars, and Venus has been observed in other planetary systems \citep{doyle+19}, suggesting that oxidizing process operating in rocky planets and their precursors is universal.

%other considerations/effects not included: 1) impacts: fragmentation, aerial bursts, mantle removal, cratering, cometray dist, comp, late veneer, 2)primordial, extended atmospheres, reduced atmospheric gases  3) final evolution, subsequent evolution

\section{Conclusions}

In this paper, we built an end-to-end model of planetary volatile growth by bringing together important processes including  impact erosion, mantle-atmosphere equilibration, and stochastic accretion.  Despite the obvious simplifications in calculating impact and exchange processes, we were able to elucidate several aspects of volatile evolution on primitive Earth-like proto-planets.
The main takeaways from this work are:

\begin{itemize}
    \item A volatile growth model with the inclusion of impact delivery and loses can roughly reproduce the atmosphere and mantle reservoir masses of present-day Earth.
    \item Impactor properties, not the primordial condition of the proto-atmosphere, determine the growth histories of volatiles.
    \item Models constrain carbonaceous chondrite water abundances to ${\sim}1-5\%$ by mass. 
    \item Models constrain planetesimal building block size distribution to one similar with the asteroid belt population. Our calculations precludes a distribution with a longer tail dominated by more massive planetesimals $>1000$ km as this assumption results in massive atmospheres in excess of 1000~bars.
    \item A suite of N-body accretion simulations, integrated with our model, highlight the wide range of potential volatile inventory for planets with Earth-similar ($0.7-1.1~M_\oplus$) masses and orbital semi-major axes. 
    \item Superchondritic C/N ratio on bulk silicate Earth is the time dependent result of impact-induced atmospheric loss and subsequent mantle degassing. Based on our model, the elevated MO C/N ratios is robust to different impact histories and initial planetesimal/embryo masses.
\end{itemize}

%primordial

\acknowledgements
H.C. acknowledge support from the Future Investigators in NASA Earth and Space Science and Technology (FINESST) Graduate Research Award 80NSSC19K1523. H.C. is supported by an appointment to the NASA Postdoctoral Program at Goddard Space Flight Center, administered by Oak Ridge Associated Universities under contract with NASA.  We acknowledge and thank the computational, storage, data analysis, and staff resources provided by the QUEST high performance computing facility at Northwestern University, which is jointly supported by the Office of the Provost, Office for Research, and Northwestern University Information Technology.

\centerline{ Supplementary Material}

\section{N-body Accretion Simulations}
To explore the effects of accretion history, we used the outputs from a suite of N-body simulations based on the Grand Tack Paradigm \citep{WalshEt2011NATURE,jacobson+morbi14} to calculate the accretion and loss history of major atmospheric constituents N$_2$, H$_2$O, and CO$_2$. The  Grand  Tack  model describes the inward and outward migration of the Jupiter and Saturn within the gas disk and is designed to reproduce many aspects of the inner solar system, for example the Martian embryo. The setup for N-body accretion simulations are similar to those described in \citet{WalshEt2011NATURE}, specifically with regards to the migration of the giant planets, radial extent of the disc, and prescription for gas drag. Each accretion scenario is different primarily by way of the initial ratios between embryos and planetesimals and the mass of the embryos themselves.  \citet{jacobson+morbi14} argued that these changes are crucial to the fully grown planetary bodies as they reflect the state of oligarchic growth; namely the maturity of the gas disc.

The Jovian planets are assumed to be nearly fully formed (with zero eccentricity) and the solar nebula has been dispersed at the beginning of the accretion of the inner planets (${\sim} 100$ Myr).
By the termination of each simulation, we deem a planet to be ``Earth-like" when it has grown to at least ${\sim} 0.7 M_\oplus$ and is situated between 0.8 and 1.2 AU.
For a more detailed description of simulations, refer to \citet{WalshEt2011NATURE} and \citet{jacobson+morbi14}. 

\section{Accretion of Major Volatile Elements}

For planetesimal accretions, we assume that all of the material mixes homogeneously into the mantle of the parent body. All starting embryos and planetesimal bulk compositions are assigned as a function of their heliocentric semi-major axis of origin. The initial volatile contents of each heliocentric distance range are derived from published cosmochemical reservoir data,
Specifically, we divide the protoplanetary disk into three distinct compositions based on chondritic meteorite type: enstatite chondrites (E-type), ordinary chondrites (S-type), carbonaceous chondrites (C-type).  %,and trans-Neptune comets\footnote{Since  originate from the same outermost part of or beyond the asteroid belt as enstatite chondrites, we do not treat them as separate populations}. 
We determined the origin heliocentric distance of each body $a$ and set its bulk composition $\chi_{\rm pl}$ based on the following step function (see also \citealt{Raymond+2004,RaymondEt2009Icar}):
\begin{equation}
\chi_{\rm pl} =\begin{cases} {\rm E{\rm -}type \hspace{2 mm} asteroids} & {\rm for} \hspace{2 mm}  a < 2 \\
                     {\rm S{\rm -}type \hspace{2 mm} asteroids}  &  {\rm for} \hspace{2 mm}   2 < a < 2.5 \\
                     {\rm C{\rm -}type \hspace{2 mm} asteroids} & {\rm for} \hspace{2 mm}   a \geq 2.5
       \end{cases}
\end{equation}

For the gas-phase species, we assume that nitrogen primarily reside in molecular nitrogen (N$_2$), carbon in carbon dioxide (CO$_2$), and H in water molecules (H$_2$O).  We assume that the carbon monoxide (CO) abundance is low because of the oxidizing conditions; for instance, the presence of photolytic water products (e.g., OH$^-$) promotes fast recombination to CO$_2$. 
Table 1 lists the adopted abundance of each gaseous species for each reservoir and their reference.
Note that many of these measured values have high uncertainty bars and we will explored this uncertainty space by varying their assumed values.
\begin{table}[htbp]
\caption{Initial Composition Inventory of Various Cosmochemical Reservoirs Expressed as Mass Fractions.}
   \label{tab:Table 1}
   \begin{center}
   \begin{tabular}{cccc}
\hline
Species & E-type & S-type & C-type\\
\hline
C & 1e-6 & 1e-4 & 0.01 \\
N$_2$ & 5e-7 & 1e-5 & 1e-4 \\
H$_2$O & 1e-5 & 0.001 & 0.05 \\
\end{tabular}
\end{center}

\footnotesize{E-type= enstatite chondrite, S-type= ordinary chondrite, C-type=  carbonaceous chondrite. Data taken from \citet{Kerridge1985GCA}, \citet{Robert1982GCA}, \citet{JessbergerEt1988}, \citet{ThomasEt1993GCA}, \citet{AbeEt2000}, and \citet{PearsonEt2006MPS}.
The uncertainty for each value is not indicated, but we will explore these uncertainties in Section~\ref{sec:results} by varying them one at a time and rerunning the model.}
\end{table}
During each accretion event, all of the projectile's mass is added to the target. While this included all of the species, we then immediately assess the effects of atmospheric impact erosion.

\section{Atmospheric Impact Erosion}

%observable fingerprints in favor of impacts 
The isotopic ratios of the silicate Earth is chondritic wheres their relative abundance are not, implying that Earth's atmosphere has experienced bulk removal via impacts rather than processes entailing isotopic fractionation such as hydrodynamic escape or mantle ingassing \citep{schlichting+18}. Indeed, impact cratering and erosion is thought to have played important roles during late stage planet formation \citep{WalkerEt1986Icar,Chyba1990NATURE,AhrensEt1993,NiemEt2012Icar}.
Impacts contribute to mass-loss by propelling large quantities of atmospheric gas into space through the formation of large vapor plumes \citep{Melosh+Vickery1989} and/or ground-motions \citep{Genda+Abe2005NATURE}.

To account for the effects of impact erosion, we follow prescription of \citet{SchlichtingEt2015Icar}, which we summarize below. 
A good approximation for the effect of an impact is as a point explosion, where the the impact's kinetic energy has been converted into heat and pressure \citep{melosh1989impact}.
This explosion pushes away atmospheric gas that could lead to significant ejection.
The precise amount of ejected mass, $m_{\rm loss}$ is dependant on the velocity and mass of the projectile.
\citet{SchlichtingEt2015Icar} identified three regimes of atmospheric mass loss dictated primarily by the size of the projectile:

\begin{equation}
m_{\rm loss} =\begin{cases} 
0 &  {\rm for} \hspace{2 mm}   r_{\rm pl} < r_{\rm min} \\
\frac{m_{\rm pl}}{M_{\rm atm}} \left[\frac{r_{\rm min}}{2 r_{\rm pl}}\left[1 - \frac{r_{\rm min}^2}{r_{\rm pl}^2}\right]\right]   &  {\rm for} \hspace{2 mm}   r_{\rm min}  < r_{\rm pl} < r_{\rm cap}  \\
\frac{M_{\rm cap}}{M_{\rm atm}} & {\rm for} \hspace{2 mm}   r_{\rm pl} \geq 2.5
       \end{cases}
\end{equation}

\noindent where $r_{\rm min}$ is the minimum planetesimal radius to eject any mass $\left(3 \rho_{\rm surf}/\rho_{\rm pl}\right)^{1/3} H_{\rm tar}$, $r_{\rm cap}$ is $\left(3 (2 \pi)^{1/2} \rho_{\rm surf}/4 \rho_{\rm pl}\right)^{1/3} \left(H_{\rm tar} R_{\rm tar}\right)^{1/2}$,  $\rho_{\rm surf}$ and $\rho_{\rm pl}$ are the surface density of the planet and the density of the planetesimal respectively, $H_{\rm tar}$ and $R$ are the scale height and the current radius of the parent proto-planet, and $M_{\rm cap}$ is the maximum amount of gas that can be ejected above the tangent plane \citep{Melosh+Vickery1989}. The scale height of the target is $(kT_{\rm a})/(\mu m g)$ where $k$ is the Boltzman's constant, $T_{\rm a}$ is the atmospheric temperature which we have assumed to be isothermal, $\mu$ is the time-evolving mean molecular weight, $m_{\rm H}$ is the mass of a hydrogen atom ($1.67\e{-24}$ g), and $g$ is the gravitational acceleration. To determine the evolving radius of the embryo $R$, we use the planetary embryo mass-radius relationship following \citet{JacobsonEt2017EPSL}.

Highly energetic collisions could also cause ground motions generating strong, potentially global-scale shocks. These shocks could propagate through a large portion of the atmosphere and induce substantial loss in atmospheric mass. For isothermal atmospheres (a good approximation for temperate $T_{\rm eq} \leq 400$ Earth-like atmospheres), the mass fraction lost by each giant impact is given by:

\begin{equation}
\chi_{loss} = 0.4 \left(Y\right) +1.4 \left(Y\right)^2 -0.8 \left(Y\right)^2
\end{equation}

\noindent where $v_{\rm imp}$ is the impact velocity, $v_{\rm esc}$ is the escape  velocity, $m$ is the mass of the projectile, $M$ is the mass of the parent proto-planet, and $Y = (v_{\rm imp} m)/(v_{\rm esc} M)$. We apply the above formalisms to calculate the mass-lost per impact.

%\section{EUV Hydrodynamic Escape}

%					Feuv=29.7*(time/1e9)**(-1.23)*(1.23)**(-2)
%					m_elim = (np.pi*Feuv*target_r**3)/(cgrav*target_m)

\section{Planetesimal Population Distribution}

The N-body simulations we use do not include true planetesimals. Rather, larger ``super-planetesimals" with minimum masses of ${\sim} 0.00251 M_\oplus$ ($1.5\e{25}$ g) are initialized in the N-body simulation. However, injecting one large planetesimal would result in different mass-lost than recurring, despite the total accreted gas amount being equal. To resolve this problem, we break down these super-planetesimals into smaller, likely more realistic planetesimals. We determine their radii by drawing, one by one, a random projectile radius from the asteroidal population distribution (e.g., \citealt{BottkeEt2005Icar}) until the total mass of the planestimals equals to the input mass of the super-planetesimal. Each smaller planetesimal is assumed to have the same heliocentric distance and thus composition as the original super-planetesimal.

\begin{table}[htbp]
\caption{Values of Henry's Law Coefficient and Constants.}
   \label{tab:Table 2}
   \begin{center}
   \begin{tabular}{ccc}
\hline
Species & $K_{\rm H, ref}$ & $C$ \\
  & ($\frac{{\rm mol}}{{\rm kg} \cdot {\rm bar}}$) &   \\
\hline
N$_2$ & 6.1E-4 & 1300 \\
CO$_2$ & 3.4E-2 & 2400 \\
H$_2$O & 7.8E-4 & 500 \\
\end{tabular}
\end{center}
Data from \citet{adamson1967physical}, measured at standard state temperature of 298 K, used to calculate different values of the Henry solubility constants.
\end{table}

\section{Mantle-Atmosphere Exchange}

Impacts between embryos and planetesimal are often directly associated with magma ocean (MO) formation, which promotes exchange between the atmosphere and underlying mantle.
To estimate the equilibration of each gaseous species between the mantle and the atmosphere after each accretion, we employ a version of Henry's law given by the following:

\begin{equation}
[A({\rm aq, i})] =  K_{\rm H, i} \left(P_{\rm tot} - h/100 P_{\rm o}\right) f_{\rm i} \label{eq:henry}
\end{equation}

\noindent where $[A({\rm aq})]$ is the concentration of gaseous phase of species i, $K_{\rm H, i}(T)$ is the Henry solubility coefficient, $P_{\rm tot}$ is the total atmospheric pressure, $h$ is the relative humidity, $P_{\rm o}$ is the vapor pressure at ambient $T$, and $f_{\rm i}$ is the mole fraction of gas i in dry air.

Pressure $P = (g M_{\rm atm})/(4 \pi R^2)$ is the surface pressure at the base of the atmosphere, where $M_{\rm atm}$ is the mass of the atmosphere and $R$ is the radius of the proto-planet, calculated from the mass-radius relationship \citep{JacobsonEt2017EPSL}.

$K_{\rm H, i}(T)$, the Henrinian solubility coefficient, is determined by the following equation:

\begin{equation}
K_{\rm H} = K_{\rm H, ref}   \hspace{1 mm} {\rm exp} \left(C \left(\frac{1}{T} - \frac{1}{T_{\rm ref}}\right)\right)
\end{equation}

\noindent where $K_{\rm H, ref}$ is the standard reference solubility constant for species $i$, $C$ is a constant in Kelvins, $T$ is the surface temperature at the base of the atmosphere which we fix at 1500 K \footnote{This assumption is valid in that the effects on composition is small, which is also noted by \cite{Pepin1997Icar}.}, and $T_{\rm ref} = 298$ is the standard state temperature. The values for $K_{\rm H, ref}$ and $C$ used are specified in Table~2. 

To find the equilibrium mass to be established in the mantle, we modify Equation~\ref{eq:henry} to the form:
\begin{equation}
m_{\rm gas, eq} =  K_{\rm H, i} P_i  M_i m_{\rm mantle}
\end{equation}

\noindent where we have replaced $\left(P_{\rm tot} - h/100 P_{\rm o}\right) f_i$ with $P_i$. $M_i$ is the molecular mass of species $i$ and $M_{\rm mantle} = 0.6 M_p$ is the current mass of the growing mantle. Whether degassing or ingassing occurs depend on both the amount of gas in the mantle and atmosphere. If $m_{\rm gas, eq}$ is greater than the current amount in the mantle, then we release the difference to the atmosphere and vice versa.

%The majority of Venus's water may have been lost prior to solidification.

\section{Core Sequestration of Carbon}

Apart from outgassing to the nascent atmosphere, core formation processes could sequester large portions of siderophiles such as carbon due to its high affinity to metals. To account for equalibration between the lower mantle and core, we use a simple partition coefficient:

\begin{equation}
D_i = \frac{C^{\rm alloy}_i}{C^{\rm sil}_i}
\end{equation}

\noindent in this equation, $C^{\rm sil}_i = M^{\rm sil}_i / m^{\rm sil}  $ and $C^{\rm alloy}_i = M^{\rm alloy}_i / m^{\rm alloy} $ is the concentration in the mantle and core respectively, where $M^{\rm sil}_i$ is the volatile component $i$ in the silicate, $m^{\rm sil}$ is the mass of the magma ocean, $ M^{\rm alloy}_i$ is the volatile component $i$ in the alloy, and $m^{\rm alloy}$ is the total mass of the core. We follow \citet{deguen2011experiments} to determine the $m$ that interacts with the metal alloy, i.e., from the expression of the melt volume:

\begin{equation}
V_m = \gamma V_{\rm imp} \frac{8\pi G \rho_p R^2_p}{3 E_m}
\end{equation}

\noindent where $\gamma = 0.15$ is a proportionality constant, $V_{\rm imp}$ is the impactor volume, $G$ is the gravitational constant, $\rho_p$ is the mean density of the proto-planet, $R_p$ is its radius, and $E_m = 9 \times 10^6$ m$^2$ s$^{-2}$ is the specific energy.  We use this equation to calculate the melt volume to the impactor volume that reacts with a given core.

\hfill \break
\hfill \break

\end{document}